\documentclass[sigconf,screen]{acmart}

\AtBeginDocument{%
  \providecommand\BibTeX{{%
    \normalfont B\kern-0.5em{\scshape i\kern-0.25em b}\kern-0.8em\TeX}}}

\setcopyright{acmcopyright}
\copyrightyear{2023}
\acmYear{2023}
\acmDOI{XXXXXXX.XXXXXXX}

\acmConference[ASPLOS '23]{28th ACM International Conference on Architectural Support for Programming Languages and Operating Systems}{March 25-29,
  2023}{Vancouver, CA}

\acmPrice{15.00}
\acmISBN{978-1-4503-XXXX-X/18/06}

\usepackage{multirow}
\usepackage{listings}
\usepackage{xcolor}

\definecolor{codegreen}{rgb}{0,0.6,0}
\definecolor{codegray}{rgb}{0.5,0.5,0.5}
\definecolor{codepurple}{rgb}{0.58,0,0.82}
\definecolor{backcolour}{rgb}{0.95,0.95,0.92}

\lstdefinestyle{ccode}{
    backgroundcolor=\color{backcolour},   
    commentstyle=\color{codegreen},
    keywordstyle=\color{magenta},
    numberstyle=\tiny\color{codegray},
    stringstyle=\color{codepurple},
    basicstyle=\ttfamily\footnotesize,
    breakatwhitespace=false,         
    breaklines=true,                 
    captionpos=b,                    
    keepspaces=true,                 
    numbers=left,                    
    numbersep=5pt,                  
    showspaces=false,                
    showstringspaces=false,
    showtabs=false,                  
    tabsize=2,
    firstnumber=1
}

\usepackage{xcolor}
\usepackage{xspace}

\usepackage{tcolorbox} % arxiv only
\newcommand{\conjecturearxiv}[2]{{\centering%
\begin{tcolorbox}[fontupper=\itshape\textsf, width=0.975\columnwidth,boxsep=0pt,top=4pt, bottom=4pt, left=10pt, right=10pt]%
\small\textbf{Conjecture #1}: #2\end{tcolorbox}\par}}

\newcommand{\conjecture}[2]{\conjecturearxiv{#1}{#2}}

\begin{document}

\title[Poking Holes in Incomplete Debug Information]{Where Did My Variable Go?\\Poking Holes in Incomplete Debug Information}


\author{Cristian Assaiante}
\email{assaiante@diag.uniroma1.it}
\orcid{0000-0001-7705-0434}
\affiliation{%
  \institution{Sapienza University of Rome}%
  %\city{Rome}
  \country{Italy}%
}

\author{Daniele Cono D'Elia}
\email{delia@diag.uniroma1.it}
\orcid{0000-0003-4358-976X}
\affiliation{%
  \institution{Sapienza University of Rome}%
  %\city{Rome}
  \country{Italy}%
}

\author{Giuseppe Antonio Di Luna}
\email{diluna@diag.uniroma1.it}
\orcid{0000-0002-7150-0972}
\affiliation{%
  \institution{Sapienza University of Rome}%
  %\city{Rome}
  \country{Italy}%
}

\author{Leo\-nardo Querzoni}% good catch CA!
\email{querzoni@diag.uniroma1.it}
\orcid{0000-0002-8711-4216}
\affiliation{%
  \institution{Sapienza University of Rome}%
  %\city{Rome}
  \country{Italy}%
}

\begin{abstract}
The availability of debug information for optimized executables can largely ease crucial tasks such as crash analysis. Source-level debuggers use this information to display program state in terms of source code, allowing users to reason on it even when optimizations alter program structure extensively. A few recent endeavors have proposed effective methodologies for identifying incorrect instances of debug information, which can mislead users by presenting them with an inconsistent program state.

In this work, we identify and study a related important problem: the completeness of debug information. Unlike correctness issues for which an unoptimized executable can serve as reference, we find there is no analogous oracle to deem when the cause behind an unreported part of program state is an unavoidable effect of optimization or a compiler implementation defect. In this scenario, we argue that empirically derived conjectures on the expected availability of debug information can serve as an effective means to expose classes of these defects.

We propose three conjectures involving variable values and study how often synthetic programs compiled with different configurations of the popular gcc and LLVM compilers deviate from them. We then discuss techniques to pinpoint the optimizations behind such violations and minimize bug reports accordingly. Our experiments revealed, among others, 24 bugs already confirmed by the developers of the gcc-gdb and clang-lldb ecosystems.
\end{abstract}




\maketitle

% !TEX root = main.tex
\section{Introduction}
\label{se:intro}
In a seminal work from four decades ago~\cite{Hennessy-TOPLAS82}, Hennessy described the emblematic conflict between the application of compiler optimizations and the ability to debug an executable program \textit{symbolically}, i.e., in terms of its source code representation. While optimizations preserve functional semantics, they can extensively alter the intermediate computations of a program, potentially leaving source-level debugging systems unable to correctly report, in terms of the original code, the current values for several such computations. The ability to debug optimized code, however, is a necessary and desirable capability~\cite{Brooks-PLDI92}.

Using unoptimized code in its place would hardly be an option, for example, if we consider core dump analysis for executables deployed in production, logic errors that emerge only in the presence of optimization (``heisenbugs'': e.g., race conditions and some memory errors~\cite{Jia-AST13,DSilva-SP15}), or programs that face heavy constraints for running time or memory usage, among others~\cite{Brooks-PLDI92}.

To cope with the above said conflict, modern compilation systems have provisions that make each code optimization cooperate in maintaining and updating debug information during executable generation~\cite{Giuseppe-ASPLOS21}. However, maintaining an accurate mapping between source-level constructs and assembly instructions is an inherently difficult task~\cite{Copperman-TOPLAS94}, leaving compiler architects with the potential to introduce bugs at each step of the process. Two recent works~\cite{Davide-PLDI20,Giuseppe-ASPLOS21} have shown how to build reliable tools that expose many of such bugs, specifically when incorrect debug information is generated. Both works share the commonality of using an unoptimized instance of a program to expose incorrect debug information in one or more optimized counterparts, building on the implicit assumption that (well-tested) unoptimized compiler settings generate an accurate reference for differential analysis.

\paragraph{The completeness problem.}
In this work, we bring our attention to a different kind of problem: identifying when an (optimized) executable contains insufficient debug information for a symbolic inspection of the state at a given program point because of implementation defects of the compiler. We will refer to it as the \textit{completeness} problem for debug information. Intuitively, in mature compilation systems, this problem affects optimized code only~\cite{Davide-PLDI20}.

Unlike correctness problems, using unoptimized compiler settings as reference is not a possibility: as we discuss in Section~\ref{se:motivation}, some optimizations may irreversibly alter the intermediate computations of a program, for instance by altering the sequence in which some source-level statements are executed, by merging or optimizing away variable instances, or by clobbering storage locations to optimize register allocation, among others. Therefore, a discrepancy in the visibility of some program construct when debugging an optimized executable versus its unoptimized counterpart does not necessarily give away an implementation defect in the compiler, since many of these effects can be prohibitive or even impossible to account for when engineering a compiler~\cite{Daniele-PLDI18,AdlTabatabai-PLDI96}.

However, we can move our attention to identifying cases when the assembly-level representation and the already-emitted debug information would allow for a symbolic inspection of specific program state elements but the compiler did not emit sufficient debug information for that. In the following, we report a code fragment from a confirmed bug that we found for gcc~\cite{bug-gcc-105161}. Our tests revealed that, even with {\tt -O1} or {\tt -Og} optimization, variable {\tt j} appears as optimized out when the program accesses the global array {\tt b}. Due to its zero value, gcc constant-folds {\tt (j)*k} to zero and {\tt j} does not need to be part of the state of the optimized program. However, optimizing compilers can model such kind of constants with specific DWARF attributes, making the value of {\tt j} available when a debugger steps on a line where {\tt j} is visible in the source. For this variable, gcc emits a DWARF debug information entry that lacks both value and location information.

% https://gcc.gnu.org/bugzilla/show_bug.cgi?id=105161
\begin{lstlisting}[language=C]%, caption=Gcc Bug Report *****, label=ex:mot]
volatile int a;
int b[10][2];
int main() {
  int i = 0, j, k;
  for (; i < 10; i++) {
    j = k = 0;
    for (; k < 1; k++)
      a = b[i][(j)*k];
  }
}
\end{lstlisting}

Interestingly, when we moved the assignment to {\tt j} before the main loop, the compiled assembly stayed the same but the variable value became visible. Our bug report saw the prompt attention of the gcc developers, with internal discussions that brought to light a scenario that the current design of gcc is unable to handle directly.

\paragraph{Our approach.} 
In this work, we present a compiler-agnostic methodology for identifying implementation defects in compilation toolchains that lead to incomplete debug information generation.

As there is no reliable oracle for differential analysis of this completeness problem, we propose to rely on empirically derived \textit{conjectures} based on the expected availability of debug information at particular program points. In particular, we identify cases where the availability of debug information for a specific program construct can be conjectured from the visibility of other program constructs that depend on it.

We present three examples of conjectures involving, respectively, call argument values, the lifetime of a live variable, and the data dependencies of a variable assignment expression. We then show how to generate synthetic programs suitable for testing one or more conjectures and how to build tools that can help pinpoint which compiler optimizations are likely behind the found violations.

We extensively test the optimization levels of several versions of the clang and gcc compilers, ultimately exposing about 38 implementation defects in their trunk version. The tests resulted in 24 already confirmed bugs: alongside clang (11) and gcc (10) bugs, some violations came from bugs in the lldb (1) and gdb (2) debuggers that we used for analyzing the respective generated executables.

We complement these experiments with a preliminary quantitative study on the availability of variable values in optimized synthetic programs, using clang and gcc releases selected quite far in time. The results suggest that developer efforts are indeed improving user debugging experience and that solutions like ours can facilitate that: for one bug fixed by the gcc developers, we measure a substantial improvement at {\tt -O1} for the studied metric that closed half of the gap with the debugger-friendly {\tt -Og}.

\paragraph{Contributions.} In summary, this paper provides:
\begin{itemize}
\item a preliminary quantitative study on debug information generated by different clang and gcc configurations;
\item three instances of conjectures for exposing classes of debug information loss due to implementation defects;
\item an open-source\footnote{The code is available at \url{https://github.com/cristianassaiante/incomplete-debuginfo}.\label{scznote}} automated pipeline for testing compilers for these conjectures and triaging any found violations;
\item an experimental evaluation of the approach on different optimization levels and versions of clang and gcc.
\end{itemize}

%% \scznote{D3bug is next!}

% !TEX root = main.tex
\section{Motivation and Current Issues}
\label{se:motivation}

In this section, we introduce basic concepts behind the \textit{completeness} problem studied in this paper and present a preliminary quantitative study on how clang and gcc versions selected quite far in time retain debug information.

\begin{figure*}[ht!]
\centering
\begin{minipage}[t]{.31\textwidth}
    \begin{minipage}[t]{1\textwidth}
        \hspace*{-0.3cm}
        \includegraphics[width=0.99\textwidth]{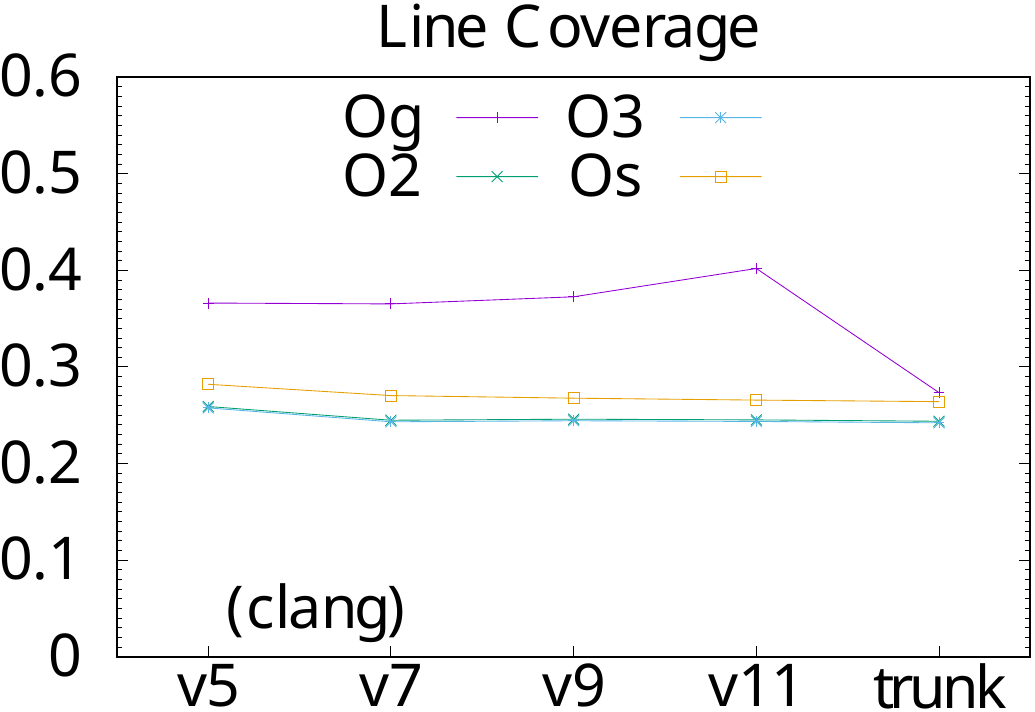}
    \end{minipage}\vspace{1pt}
    \begin{minipage}[t]{1\textwidth}
        \hspace*{-0.3cm}
        \includegraphics[width=0.99\textwidth]{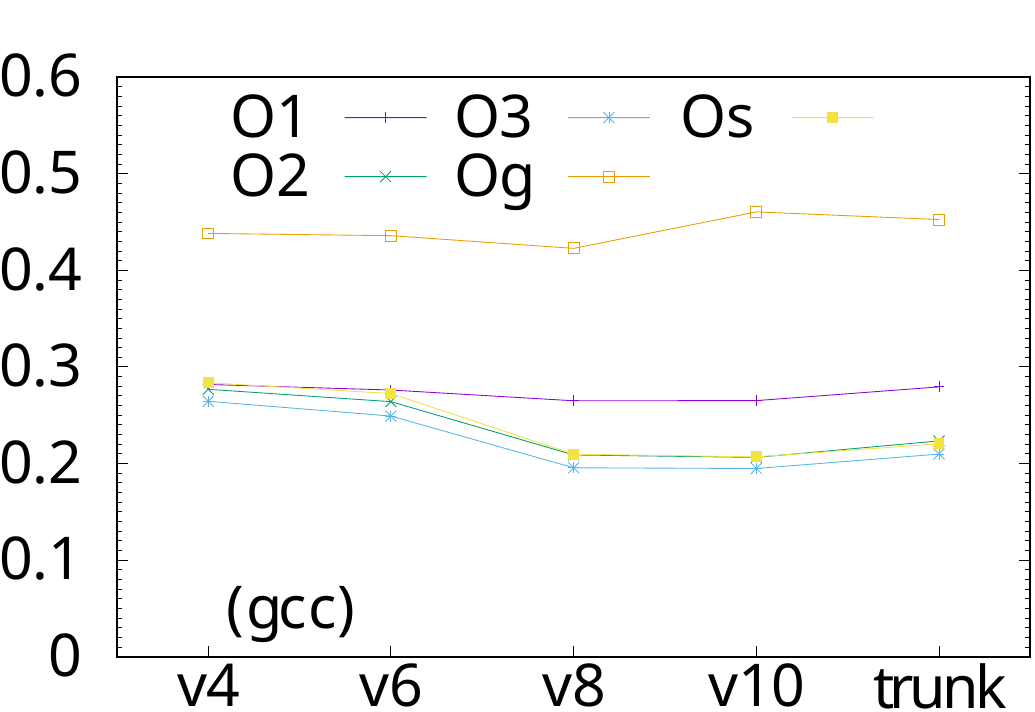}
    \end{minipage}
\end{minipage}%
\begin{minipage}[t]{.31\textwidth}
    \begin{minipage}[t]{1\textwidth}
        \hspace*{-0.15cm}
        \includegraphics[width=0.99\textwidth]{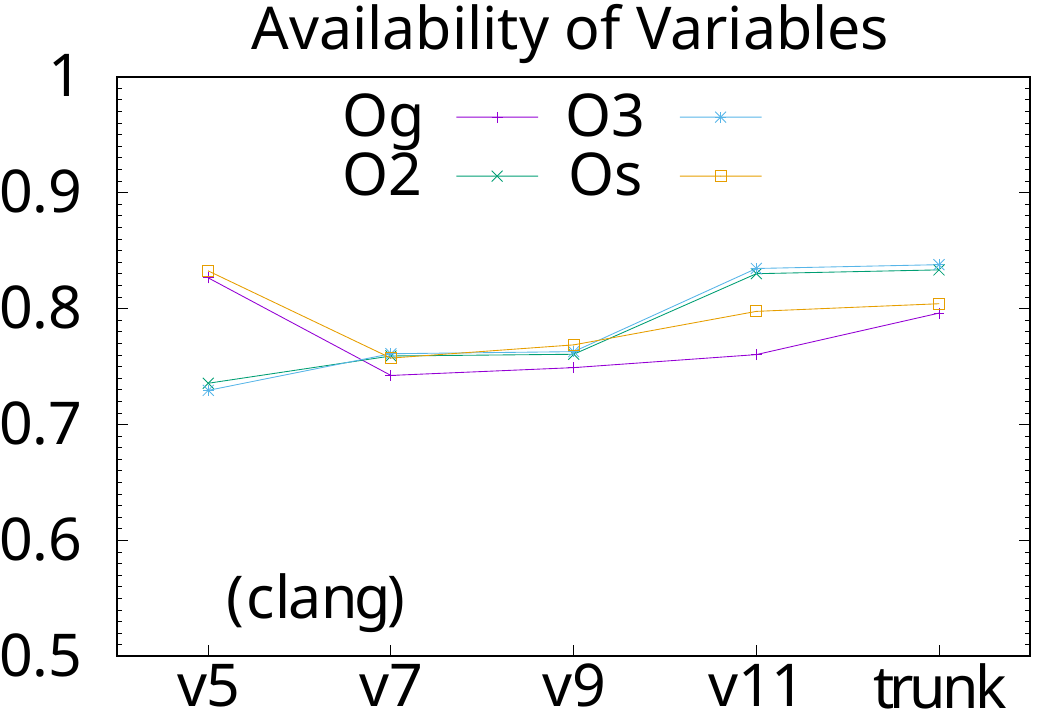}
    \end{minipage}\vspace{1pt}
    \begin{minipage}[t]{1\textwidth}
        \hspace*{-0.15cm}
        \includegraphics[width=0.99\textwidth]{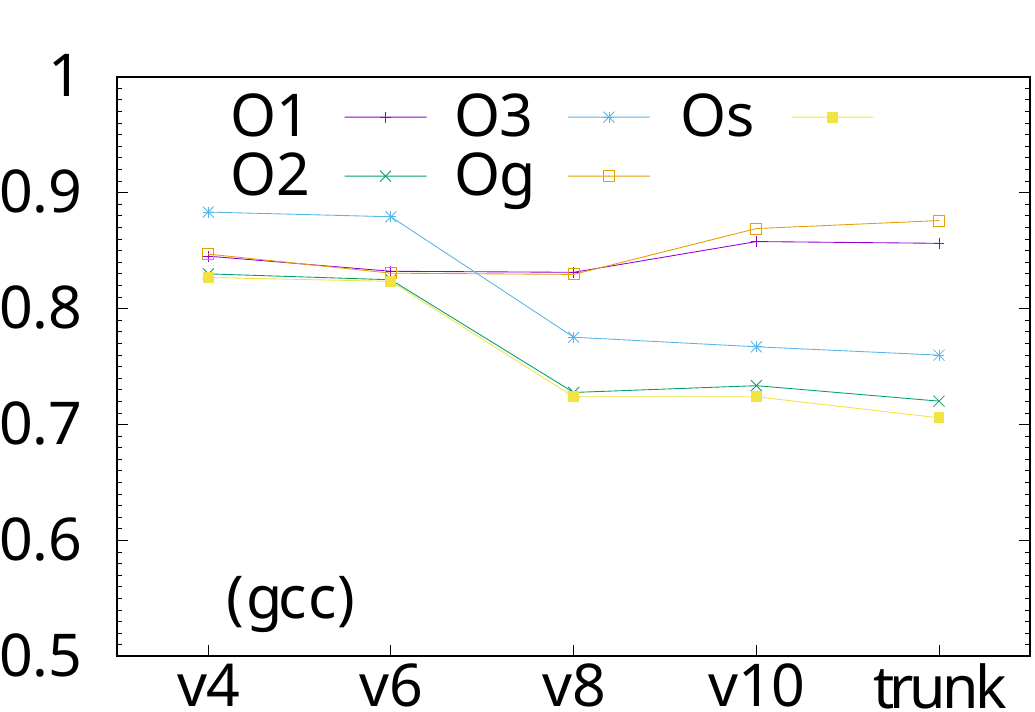}
    \end{minipage}
\end{minipage}%
\begin{minipage}[t]{.31\textwidth}
    \begin{minipage}[t]{1\textwidth}
        \includegraphics[width=0.99\textwidth]{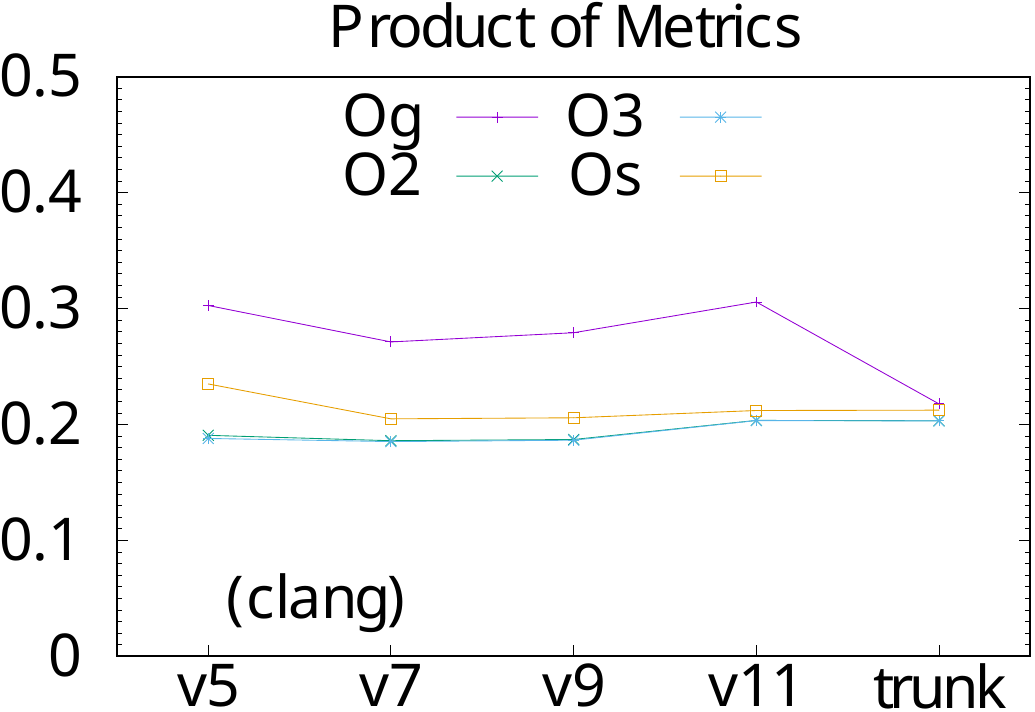}
    \end{minipage}\vspace{1pt}
    \begin{minipage}[t]{1\textwidth}
        \includegraphics[width=0.99\textwidth]{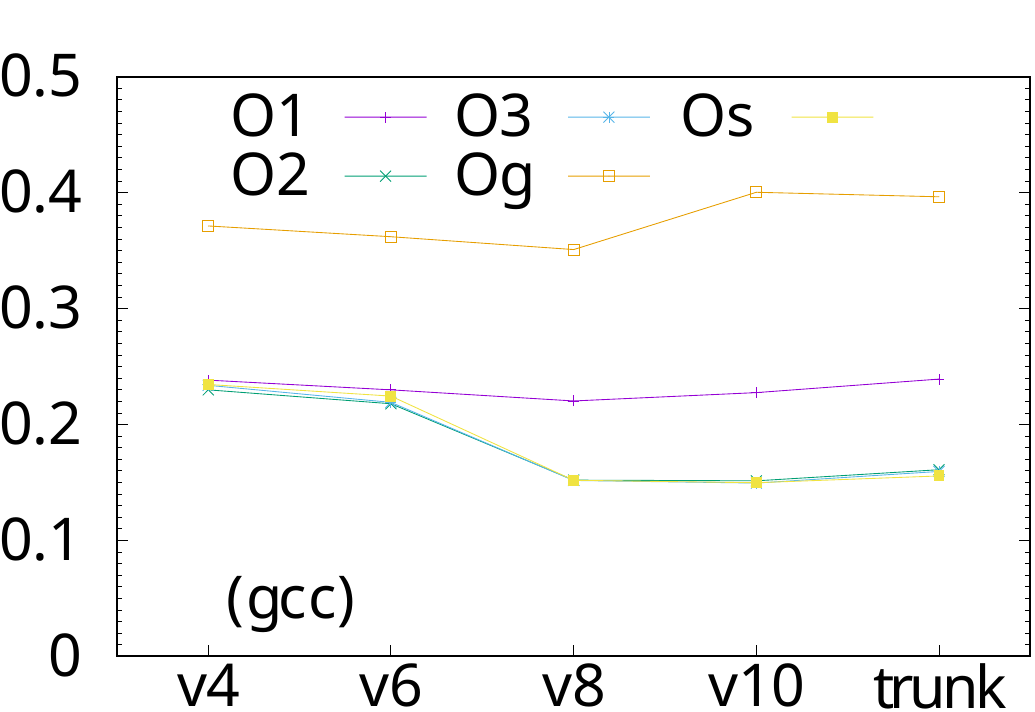}
    \end{minipage}
\end{minipage}
\caption{Statistics on debug information collected for 5K testing programs over different compiler versions.\label{fig:motivation}}
\end{figure*}

\paragraph{Preliminaries.} Optimizing compilers can deeply alter the intermediate computations of a program portion, preserving semantic equivalence only for its externally visible effects (e.g., their outputs on completion). Intuitively, this conflicts with the desirable ability to debug an optimized program in terms of its source representation.

% https://blog.tartanllama.xyz/writing-a-linux-debugger-elf-dwarf/
Compilation systems try to track and account for such changes by maintaining and updating debug information during executable generation. In the UNIX realm, debug information is eventually attached to the output executable using dedicated sections. Some compilers may produce instead a separate file, as with Visual Studio on Windows. DWARF is a general debug information format that is most commonly used (but not exclusively) with ELF files.

Formats like DWARF allow an optimizing compiler to instruct a source-level debugger tool on how the source code relates to instructions and data of the executable at hand. Therefore, these debuggers become able to display, among others, if the current assembly instruction corresponds to some source line and what are the values of the source-level variables visible in the current scope.

% \looseness=-1
\paragraph{Terminology.} In the remainder of the paper, we say that a debugger can \textbf{step} on a source line when debug information contains location information for one or more (e.g., think of loop unrolling) assembly instructions mapped to that line. We say that a variable is \textbf{visible} at a line when a debugger can step on the line and the local frame shown by debugger information includes the variable. We say that a variable is \textbf{available} at a line when it is visible and its current value can be displayed. In fact, the value of a variable may have been clobbered by optimizations (\textit{optimized-out} case) or, more generally, debug information reports the variable but misses its current location.

\paragraph{Incomplete program state.} An inevitable effect of optimization is that a debugger may often present the user only with a \textit{partial} view of the program state, compared to what the user would expect by looking at the source code. Generally speaking, some heavy-duty optimizations significantly alter the working of a program, changing the order of statements or altering the values of variables.

Some of these changes would require explicit logging or compensation machinery to undo the effects of optimization and allow for a faithful representation of the program state~\cite{Daniele-PLDI18}. The run-time overheads that logging would introduce and the prohibitive complexity behind engineering a compensation machinery lead modern compilers not to follow this path. However, there are also changes whose effects would not require either approach to be accounted for, but compilers fail to capture them because of \textit{implementation defects}, leading to an incomplete presentation of program state when debugging.

The reasons behind such defects may be different: from public developer discussions following on our reports, we can name: unanticipated interactions between multiple optimizations, lack of internal design provisions for specific patterns, absence in the current DWARF specification of constructs to capture specific patterns in a straightforward manner, and regressions induced by handling code added for other patterns. Such defects mainly arise in the compilation systems, but we found also cases where their reference debugger tools process the emitted debug information incorrectly. As there are currently no automated testing methodologies to expose issues of this kind, their identification currently hinges mainly on a ``proactive awareness'' of compiler architects, when developing an optimization, of how and where such issues may occur.

%As modern optimization pipelines are very complex, designing automated testing methods for debug information completeness bugs [...] pointing implementation defects out can [...] % bring timely enhancements and ease the work of developers.

\paragraph{Quantitative study.} We conducted a preliminary study on the debug information that different versions of the two most popular compilation systems, namely clang and gcc, generate for a pool of testing programs at different optimization levels. We generate 5000 subjects\footnotemark{} with a compiler fuzzing tool~\cite{Yang-PLDI11} and measure in a debugger how many source-code lines can be stepped on and how many variables are visible at each stepped line. 

\footnotetext{The metrics we study here reached a plateau around this pool size.}

We consider as optimization levels {\tt -O1}, {\tt -O2}, {\tt -O3}, the ``debugger-friendly'' {\tt -Og}, and {\tt -Os} that reduces program size. We leave out {\tt -Oz} as not all versions support it. For clang only, {\tt -O1} and {\tt -Og} are aliases for the optimizer, therefore we report data only for {\tt -Og}. We select versions quite far in time: we choose releases 4.8, 6.5, 8.4, and 10.3 plus trunk version {\tt 500d3f0} for gcc and releases 5.0, 7.0, 9.0, and 11.1 plus trunk version {\tt c2c977c} for clang. We compile the code for the x86\_64 architecture and use gdb 11.2 and lldb 13.0 to study the programs dynamically.

Compared to a static DWARF information inspection, testing in a debugger lets us remove noise effects from unreachable code and avoid reimplementing the (normally) well-tested native logic of debuggers when parsing DWARF information for line and variable visibility. We compile each program also at {-\tt O0} in the respective compiler version and compute these two metrics (as global average) by debugging each optimized executable instance:
\begin{itemize}
\item \textbf{line coverage}: the ratio of unique source lines that the debugger can step on compared to its {\tt -O0} counterpart;
\item \textbf{availability of variables}: the average ratio of available variables (i.e., shown with value) compared to its {\tt -O0} counterpart for the individual source lines that can be stepped on in both program instances when debugging.
\end{itemize}

% We instrument each program with a one-time breakpoint on each source line and record at each location what variables are visible and if their value is available.

\noindent
These two metrics capture altogether the two kinds of debug information loss that are possible: those that are inescapable consequences of optimization and those that come from implementation defects. As we mentioned earlier in the paper, there is currently no oracle to reliably determine to which category a lost line or variable belongs to. However, these two metrics allow us to showcase the gaps existing between optimization levels and how these gaps evolve along with compiler development. Especially when we combine them, we can speculate that most improvements come from optimizer enhancements that lead to more program state being tracked. Figure~\ref{fig:motivation} shows the results collected for all versions and optimization levels.

If we look at \textbf{line coverage}, we can see that {\tt -Og} preserves significantly more lines than any other level for both clang and gcc. This is true also across versions, with the exception of the latest versions of clang. By manual analysis of their compiled programs, we found that clang recently enabled more aggressive optimizations that avoid generating code for some loops already at {\tt -O1}/{\tt -Og}. For clang, we also observe that the size reduction heuristics of {\tt -Os} indirectly favor the possibility of stepping on more lines, since preventing some inlining or unrolling choices denies later optimization opportunities that, in turn, cause more debug information loss. Finally, we notice that gcc at {\tt -O3} ``drops'' more lines than {\tt -O2} (whereas with clang they yield nearly identical results) and that their gap with {\tt -O1} widens when comparing the 6.5 and 8.4 releases.

%\looseness=-1
If we look at \textbf{availability of variables}, we spot interesting trends on the optimization levels of each compiler. Beware this metric does not account for line visibility, hence it should not be used instead to compare different optimization levels, as they entail very different line ratios (e.g., if a level covers many fewer lines, a higher availability-of-variables value may be misleading). For clang, we observe an apparent regression between major releases 5.0 and 7.0 for {\tt -Og} and {\tt -Os}, which we speculate comes from aggressive transformations added to the pipeline; then, all optimization levels see their values increase following release 7.0, especially the most aggressive ones. For gcc, since release 8.0 we observe that the results for {\tt -O1} and {\tt -Og} tend to improve, while the other levels see a regression on that specific version (for 8.0 we can observe an analogous regression also for the other metric).

To compare the availability of variables between optimization levels, we can factor in the line coverage and compute their product. For gcc, it now becomes apparent how {\tt -Og} and, to a good extent, {\tt -O1} retain debug information for significantly more variables per stepped program point. For clang, we observe similar trends until the latest releases which, as we mentioned above, enabled even for {\tt -Og} more aggressive optimizations that remove code.

\paragraph{Takeaways.}
The improvements across releases for individual optimization levels appear indicative of the efforts that developers try to put in preserving increasingly more debug information in the operation of an optimizer. When new optimizations are added to a level, though, a regression may be inevitable. Looking at the combined product metric, gcc seems to preserve substantially more information than clang at the {\tt -Og} and {\tt -O1} levels, while the difference is modest at other levels. We also learned that different compilation systems may make different choices on retaining debug information (or even generating assembly code) for different source lines. We hope the preliminary evidence we collected can foster in-depth studies, for example, on the effects of individual transformations introduced or enabled across compiler releases.

On a different note, while one can test how later compiler versions generate supposedly faster code for each program, the same may not be done for debug information. Our metrics only describe how a compiler version fares compared to the (virtually unattainable) results of the unoptimized reference. Unfortunately, the state of the art offers no empirical means to test, in ever-evolving compilation systems, when some debug information becomes unavailable as an inevitable consequence of an introduced optimization or because of newly added or latent implementation defects.

\textit{However, identifying and fixing such defects can even have a positive ripple effect.} Later in the paper, we discuss a bug that we identified in gcc with our methodology: as the component impacted multiple optimizations, fixing it closed half of the gap between {\tt -O1} and {\tt -Og} in the availability-of-variables metric.

% !TEX root = main.tex
\section{Proposed Approach}
\label{se:approach}

In this section, we outline our proposal for identifying implementation defects in optimizing compilers behind incomplete debug information. After discussing why a conjecture-based approach can be effective for this task and what properties such a conjecture should have, we propose three possible conjecture embodiments, also detailing examples of confirmed bugs we found with them.

\subsection{Rationale and Desired Properties}
\label{ss:invariant-intro}

The idea of checking for properties, invariants, and similars at program points with specific characteristics has been successfully explored over the years in several software testing scenarios (e.g.,~\cite{Schuler-ISSTA09, Ernst-ICSE99,Fioraldi-USENIX21}). In recent efforts on testing debug information~\cite{Davide-PLDI20,Giuseppe-ASPLOS21}, an unoptimized program instance serves as reference for checking the correctness of presented values and other stack frame elements.

Unfortunately, for the many reasons we discussed in Section~\ref{se:motivation}, at the moment there is instead no reliable oracle that may tell when (or where) a given piece of program state should be visible when debugging an optimized executable instance of that program.

However, for specific code constructs and patterns, the expected presence of debug information may become predictable as we factor in reasoning and experience about what compilers can or cannot do over them.
For example, we know that an optimizing compiler cannot alter the values for arguments involved in a call to an external function, otherwise program semantics could be affected. Therefore, if a program variable appears as a call argument to such a function, one would expect debug information to correctly track its value where the call happens, making the variable available.

%\looseness=-1
\textbf{Conjectures} of this kind can be drawn from experience in compiler construction and practical observations. As with likely invariants~\cite{Sahoo-ASPLOS13}, we consider them \textit{empirically-derived conjectures}; in the remainder of the paper, we will often refer to them simply as ``conjectures'' for brevity.

To be used in systematic compiler testing, an effective type of conjecture may aim for the following properties:
\begin{itemize}
\item be verifiable in terms of source-language constructs;
\item rely on general compiler construction concepts;
\item avoid making assumptions on specific sequences of optimizations that are being applied to a program.
\end{itemize}

\noindent
These properties would allow embodiments of our approach to build on existing tools (e.g., using source-level debuggers for dynamic checking of conjectures) and to benefit from synergies with well-tested methodologies (e.g., to synthesize testing code that stresses optimizers).

In this work, we opt for \textit{compiler-agnostic} conjectures. As systematic testing for incomplete debug information is, in effect, an unprecedented task, this choice may help us expose (possibly long-standing) issues in multiple compilers without risks of overfitting the approach around the working of a specific optimizer. The three conjectures we present next turned out to be very effective in practice in exposing implementation defects on the two most popular C/C++ compilation systems to date (Section~\ref{se:evaluation}).  

The focus of our conjectures will be \textbf{available variables}, specifically when compiler implementation defects lead a variable to inadvertently appear as optimized-out or to not be reported at all in the current frame during debugging. We prioritize this dimension as it intuitively impacts user debugging experience by presenting them with a partial representation of the true program state.

\subsection{Conjecture 1: Visibility of Call Argument Sources}
\label{ss:invariant-one}

To present our first conjecture, we use as running example a confirmed bug that we reported for the {\tt InstructionCombining} peephole optimization of clang~\cite{bug-clang-49975}:

\begin{lstlisting}[language=C]%, caption=Clang Bug Report *****, label=ex:c1]
void foo(int, int, int, int, int, int, int);
static short a = 4;
void b(int c) {
    short v1 = 0;
    int v2,  v3 = 2,  v4 = 9,  v5 = 5,
        v6 = 5, v7 = (v2 = a) == 0 & c;
    foo(v1, v2, v3, v4, v5, v6, v7);
}
int main () {
    b(a);
    a = 0;
}
\end{lstlisting}

In this example, when debugging the binary that gcc generates with {\tt -O3} optimization, variable {\tt v2} does not appear among the variables visible in the frame at the call to the external function {\tt foo}. However, the optimizer is aware of the variable's use when the code passes it by copy to {\tt foo}, as it emits code to materialize the associated value ({\tt 4} from the assignment {\tt v2=a}) as argument for the call: we consider this an implementation defect. The developers identified a loss of debug information that could be avoided when simplifying the {\tt \&} operation. In general, we can identify violations of this kind by checking for:

% variable vs variable instance

\conjecture{1}{When a program variable appears as an argument for a call to an opaque function, the variable should be visible along with its value when stepping on the source line containing the call.}

By \textit{opaque function}, we mean that the optimizer does not have knowledge of the target and its effects: therefore, it cannot optimize away the variable or alter its value as a result of inter-procedural analyses~\cite{Daniele-PLDI18, Davide-PLDI20}. For instance, if a loop induction variable is used as argument, the optimizer cannot canonicalize it or reverse the loop, as the semantics of the program may be altered (e.g., the target function may use it to index volatile memory). A function defined in a different compilation unit is the most obvious example of opaque function, but other constructs are also possible: for example, think of indirect calls with varying targets.

In the defects we reported, we found several optimizations that led a variable to be entirely missing from the shown frame (like above), but also others that cause it to appear as optimized out; the first scenario was prevalent.

\subsection{Conjecture 2: Availability of Constituents}
\label{ss:invariant-three}

For our second conjecture, we discuss a program that exposed two related bugs~\cite{bug-clang-53855} in the {\tt LoopStrengthReduce} (LSR) optimization of clang, which optimizes uses of loop induction variables:

\begin{lstlisting}[language=C]%, caption=Clang Bug Report *****, label=ex:c2]
volatile unsigned int c = 0;
int a[2][4][4] = {{{1, 2, 3, 4}, ...}, ...};
unsigned short b[4] = {1, 2, 3, 4};
int main (void) {
    int i, j, k;
    for (i = 0; i < 2; i++)
        for (j = 0; j < 4; j++)
            for (k = 0; k < 4; k++)
                c = a[i][j][k];
    for (i = 0; i < 4; i++)
        c = b[i];
    return 0;
}
\end{lstlisting}

\looseness=-1
The program writes to a volatile global variable and reads from two global arrays, using one or more loop induction variables to index such arrays. The nature of memory here restricts the work of optimizers, as they have to preserve the visibility of each update. Variable {\tt i} operates as the induction variable for the outermost loop and for the subsequent loop. When stepping on the two lines that update variable {\tt c}, we found that {\tt i} is visible at both with {\tt -O2/-O3}, only at the second with {\tt -O1/Og/Oz}, and only at the first with {\tt -Os}. The developers confirmed that the provisions that LSR has to ``salvage'' (in clang jargon) debug information were insufficient for this program.

Our second conjecture involves the data dependencies of an assignment: we reason on what variables we expect to be available with their value when stepping on the source line of the assignment.

In general, though, the attentive reader may argue that for an expression like {\tt v1=v2+v3*v4}, an optimizer is free to generate code that, for instance, computes the result by reusing the storage location of one or more expression constituents if the reminder of the program does not use them (i.e., a variable become dead). The absence of debug information to display their value may thus be just an inevitable effect of optimization. Therefore, we propose to focus on assignments with specific properties:

\conjecture{2}{When stepping on a source-code line that assigns a value to global storage through a non-simplifiable expression, we expect a variable $x$ taking part in the value computation to be visible at that line if (i) $x$ is a constant or (ii) optimizations cannot alter the value of $x$ and the program may use $x$ later.}

The conjecture comes with three choices. First, we rule out trivially simplifiable expressions, such as {\tt v1=v2\&0}, where not all constituents are necessary for the result.

\looseness=-1
Second, we focus on lines that assign to global storage: when a debugger can step on one, it will happen on instructions that are about to make the change visible externally. This may not be the case for local variables, for which the value computation may not take place (we discuss such a case for Conjecture 3) or may not reflect the source-level semantics due to optimizations that alter the value (then a compiler may even opt for hiding it~\cite{LeChun-PLDI99}).

Third, for the variables taking part in the value computation, we check the visibility only of those that either hold constant values (therefore trivial to track in debug information) or that the optimizer should leave untouched (e.g., loop induction variables used to reference a location from global memory as in the example above).

For variables of the first type, by \textit{constant} we include variables assigned with a numeric or string literal or that take the address of another variable. For variables of the second kind, we can check their downstream uses (e.g., liveness~\cite{Daniele-PLDI18}) in the program to be confident that an optimizer cannot reuse their storage during value computation (or we could be dealing with valid optimized-out cases).

\subsection{Conjecture 3: Decaying Visibility of a Variable}
\label{ss:invariant-two}

For our last conjecture, we discuss a confirmed gcc bug~\cite{bug-gcc-104938} for its tree-based sparse conditional constant propagation ({\tt -ftree-ccp}):

\begin{lstlisting}[language=C]%, caption=Gcc Bug Report *****, label=ex:c3]
char a = 0;
int b = 0;
void foo(int *d) { a = 0; }
int main() {
    int *v1 = &b;
    int **v2 = &v1;
f:  if (a)
        goto f;
    *v2 = v1;
    foo(*v2);
}
\end{lstlisting}

Interestingly, the bug surfaces only when compiling the code with the debugger-friendly {\tt -Og} optimization level, while the missing variable value is available at more aggressive levels. The bug involves variable {\tt v1}: its value is displayed as optimized-out since its assignment, only to become visible when reaching the function call to {\tt foo}. This behavior is counter-intuitive: the visibility of a variable is expected to only degrade in the remainder of its lifetime (e.g., from available to optimized-out) because the optimizer may claim for its storage, if any.

The analysis of the bug revealed that all optimization levels lead {\tt main} to no longer have instructions that assign the two variables. In optimization levels other than {\tt -Og}, the first assembly instruction for {\tt main} becomes a load from global variable {\tt a} that is hoisted out of the {\tt if-goto} loop, whereas {\tt -Og} does not host it and the emitted DWARF range information for {\tt v1} make its value available only well after the loop. In general, we can look for violations of this kind by checking for the following:

\conjecture{3}{When a function assigns to a local variable and a subsequent source line can be stepped on, the availability of the variable value can only remain the same or worsen in the remainder of the program.}

For this conjecture, special attention can be devoted to reassignments of a variable value at different program locations, which are the only behavior allowed to ``refresh'' the visibility of a variable and can be treated as different variable instances.

\subsection{Discussion}
\label{ss:discussion}

The three provided examples of conjectures are meant to capture recurrent patterns and constructs in the code generation and debugging practice in ways that are amenable to automated verification and simple to reason about.

As presented, the conjectures came from progressive refinements over key ideas originated from a mix of intuition and experience. The refinements involved restricting the classes of constituents on which they hold, often based on what we saw in the assembly (i.e., the optimizer kept the values in the machine-level state) during preliminary experiments. We speculate this task should be easier for compiler developers due to their knowledge of optimization designs, which may also favor the identification of new properties.

Being empirically derived, the soundness of a conjecture can only be argued for by drawing from theoretical arguments from compiler construction, analytical and statistical observations on debugger traces, and, ultimately, by the feedback that compiler architects and developers provide for the reported violations. However imperfect that may sound, we believe that a testing solution for this kind is, for the time being, an effective and unprecedented way to cope with the lack of any oracle for systematic differential testing. The empirical evidence we collect (Section~\ref{se:evaluation}) supports this argument, as we are able to expose bugs that involve heterogeneous compilers and their optimizations.

Moving away from the properties of Section~\ref{ss:invariant-intro}---for instance, by reasoning on the intermediate representation of a compiler---may be convenient to expose corner cases in the implementation of a compiler. However, it may require significant expertise in a specific toolchain and limit reusability across compilation systems. On the contrary, new compiler-agnostic conjectures could be explored in future work, even to test multiple debugging dimensions.

On a different note, the attentive reader may point out that both Conjecture 2 and 3 relate to liveness properties of variables to some extent. For Conjecture 2, liveness is a shortcut that can be used to avoid false positives when the optimizer reuses the storage of an ``unalterable'' variable for computing the assignment under analysis.  For Conjecture 3, one cannot rely (only) on source-level liveness analysis as an optimizer may move around statements that do not depend on the variable under analysis and, more importantly, for non-constant variables the value may remain visible in the debugger even when its lifetime ends (until the optimizer claims the storage). In reality, Conjecture 3 reasons on the lifetime of a single variable in terms of its visibility in a debugging session.

% !TEX root = main.tex
\section{Identifying and Understanding Violations}
\label{se:violations}

%\looseness=-1
This section details how we can obtain programs for testing a compilation system against the conjectures of Section~\ref{se:approach}, pinpoint which optimizer components are likely behind a violation, and minimize the testing program to ease analysis by developers. {The components we developed to this end are publicly available (link in footnote~\ref{scznote} of Section~\ref{se:intro}); some may be of independent interest. Overall, they are made of \textasciitilde250 Bash/Perl and \textasciitilde1700 Python LOC.

\subsection{Test Subject Generation}
\label{ss:testcase-generation}

Due to the extensive code bases of modern compilation systems, to spot completeness bugs we seek for sufficiently heterogeneous test cases that undergo different and multiple optimizations, so that debug information is affected by their combined effects.

% An empirical comparison of compiler testing techniques ICSE 16 https://dl.acm.org/doi/10.1145/2884781.2884878
The development of optimizing compilers has benefited over the years from supporting tools such as regression and torture-test suites to point out code generation bugs. Lately, generative fuzzing-style testing tools have been particularly successful in exposing compiler bugs, as reflected by recent proposals (e.g.,~\cite{CompBugs-OOPSLA16, CompBugs-ISSTA18}) and studies (e.g.,~\cite{CompBugs-PLDI13, CompBugs-ICSE16}) on correctness testing for code generation. 

In light of the proven efficacy of such tools in exercising heterogeneous behaviors of an optimizer, we speculate they can be similarly useful also in introducing (both inadvertent and unavoidable) losses of debug information during compilation. An analogous speculation proved effective in debug information correctness testing~\cite{Giuseppe-ASPLOS21}.

We work with the popular Csmith fuzzer~\cite{Yang-PLDI11} to generate programs to be checked for our conjectures. We configure it to draw every time from different assortments of 20 options that define program characteristics. We then reuse identical programs to test the three conjectures.

Albeit writing special-purpose code generators for single conjectures may be lucrative in some cases, we believe this would contrast with the generality that we pursue for compilation constructs (Section~\ref{ss:invariant-intro}), which in turn mitigates the risk of inadvertently restricting the pool of optimizations which an optimizer may draw from.

\subsection{Conjecture Violation Checking}
\label{ss:checking-conjectures}
We check if a conjecture is violated at a program point by means of dynamic analysis in one or more debugging tools. For an optimized executable obtained with a given compiler, we use the native debugger tool for that compiler (lldb for clang, gdb for gcc) to check whether the expected variable(s) are visible on the currently analyzed line. This choice lets us test the optimizer and the reference debugger of a compilation system simultaneously.

% %\footnote{Vice versa, a ``non-native'' debugger may fail to display the value for a program state element due to differences in how another compiler generates debug information compared to how the debugger expects it.}.

When a violation is met, we repeat the test also in a different debugger and in other versions of the same debugger, to expose cases where complete debug information is present but some bug affects the native debugger.
During test generation and, more extensively, when validating a violation, we also check whether the program exhibits undefined behavior using standard tools (i.e., compile-time checks eventually followed by static analysis in {\tt compcert}~\cite{CompCert-Com09}).

For the three conjectures presented in this paper, we find it typically sufficient to check them only the first time a source line is met during debugging\footnotemark{}. Therefore, we inspect executables using standard tools (e.g., {\tt readelf}) to extract what source lines can be stepped on and run the program in the debugger, configuring the latter to place a one-time breakpoint on each such line. We then record a simple trace where, for each line covered by the execution, we save the identity of the variables visible in the frame of the function and their displayed value, if any.

For Conjecture 1, we cannot influence the arguments and destinations that Csmith chooses for function calls. Thus, we modify and recompile its programs by linking an external code module containing a non-optimizable function~\cite{Davide-PLDI20} (i.e., a stub making a {\tt printf} operation on its arguments) and adding a call to it at a random source line, choosing as arguments for the call a plurality of the local variables. In the debugger trace, we then check if the respective variable values are visible at the call. Conjecture 2 and 3 do not require program modifications, therefore we generate a single debugger trace for analysis.

\subsection{Looking for the Culprit Optimization}
\label{ss:multi-scz}
Depending on the abundance of implementation defects in a system and the characteristics of the testing programs, current tools for compiler testing can produce high numbers of tests that require prioritization techniques for subsequent analyses. This ``compiler-fuzzer taming problem''~\cite{CompBugs-PLDI13} occurs also with completeness issues from violated conjectures, as we measure in Section~\ref{se:evaluation}.

\footnotetext{Optimizations like loop unrolling may introduce multiple instances of the same source line with different completeness properties. However, checking loops in full can be very time-consuming. During exhaustive early tests, we found similar cases only when the first loop iteration is peeled---a scenario captured by the proposed criterion.}

A recent endeavor on correctness testing of debug information~\cite{Giuseppe-ASPLOS21} proposes to use the bisection method of clang, which makes the work of the optimizer's pipeline stop after a controlled number of iterations and allows for a differential analysis of the executable. This makes it possible to determine what is the optimization transformation that, once applied, makes the information loss visible\footnote{Sometimes the root cause may also be an optimization applied early. However, in a black-box testing, this is the only viable option, leaving the identification of the root cause to manual analysis.}. We adopt this technique for grouping violations exposed by a specific conjecture on clang-optimized programs.

Unfortunately, the method is not applicable to compilation systems like gcc that cannot be configured to work incrementally (at least, not without tweaking its internals). Therefore, we propose a simple solution that may be of independent interest. We surveyed the compilation options for gcc implied by each optimization level and collected all the boolean flags {\tt -fno-opt} that restrict optimization. Given a program and the optimization level at which the violation was found, we recompile the program for verification by indicating the same optimization level and one of the {\tt -fno-opt} flags, trying each of them separately to see if the violation no longer occurs. The number of flags to test was 81-151 depending on the optimization level.

%\
Due to dependencies between optimizations (for example, turning off inlining prevents other optimizations from happening), our method sometimes identifies multiple flags. Those can be analyzed in further combinatorial assortments or be heuristically prioritized according to experience (as with the inlining example, by giving inlining-related options a low rank). The method fails only when a behavior cannot be controlled by flags (as with some {\tt -Og} internals) or when more than one optimization should be disabled to make the violation no longer occur.

\subsection{Minimizing a Test Program}
A downside of using a generic generational approach to generate test programs is that the output often consist of hundreds of lines of code. In particular, the settings that we profitably used for Csmith typically led to 400-500 lines in the C language. Filing lengthy code in bug reports may take away precious time from (and, unfortunately, discourage) developers in doing a prompt inspection of the internal work of the optimizer to understand the issue.

We build on C-Reduce, a state-of-the-art solution~\cite{CReduce-PLDI12} to test case reduction for bugs in code generation, and augment it with machinery to preserve the conjecture violation that we identified with the techniques of the previous sections. Compared to the correctness testing work of~\cite{Giuseppe-ASPLOS21}, we add provisions to preserve the identified culprit optimization, as we observed that the extensive changes made by C-Reduce can lead to reduced programs where the conjecture is violated at the same line but a different optimization is behind it. Preserving the culprit optimization is important as it maintains the soundness of the by-group prioritization criterion for bug reporting (Section~\ref{ss:multi-scz}) and prevents a more ``dominant'' buggy optimization to mask other issues in the optimizer.

At each reduction step, we compile the program two times: one with the optimization level that made the violation emerge and one where we also disable the culprit optimization. If the reduction step preserves the culprit optimization, the violation will not occur in the second program and the reduction can be accepted.

Finally, we extract the assembly code and the relevant debug information (e.g., the DWARF DIE data for the involved variables) for the optimized reduced program compiled with and without the identified flag. In our experience, the differences between the two versions have proven helpful to ease and prioritize bug analysis, especially when the assembly code resulted as unchanged.

% !TEX root = main.tex

\section{Experimental Results}
\label{se:evaluation}

%We evaluate our proposal in the following respects:
In this section, we discuss the experimental findings that we collected by applying our approach to multiple configurations of the clang and gcc compilers. In particular:
\begin{enumerate}%[label=RQ\arabic*]
\item we study how often recent compiler versions generate code that violates any of our three conjectures;
\item we investigate whether the conjectures can expose defects in heterogeneous components of an optimizer;
\item we describe both common and peculiar traits of the bugs that we reported to compiler developers;
\item extending the study of Section~\ref{se:motivation}, we analyze a selection of compiler versions retrospectively.
\end{enumerate}

\paragraph{Methodology.}
We run our tests on a server equipped with an Intel Xeon E5-2699 CPU, 256 GB of RAM, Linux OpenNebula3, kernel 4.4.0, with modest background activity. We generate 1000 test programs and use them to check the three conjectures across different compiler configurations. Details on the used compiler versions and optimization levels are provided in the next sections. As reference debugger tools, we use gdb 11.2 and lldb 13.0 (latest stable versions) as done for the study of Section~\ref{se:motivation}.

\subsection{Violations in Latest Compiler Versions}
\label{ss:eval-violations-trunk}

\looseness=-1
As the first dimension of our study, we tested the latest trunk versions at evaluation time---{\tt 500d3f0} for gcc and {\tt c2c977c} for clang---against our conjectures, generating executables for an x86\_64 machine at optimization levels {\tt -Og}, {\tt -O1}, {\tt -O2}, {\tt -O3}, {\tt -Os}, and {\tt -Oz}. As {\tt -O1} and {\tt -Og} are currently identical in clang, we report only {\tt -Og} for it.

Table~\ref{tab:gcc-trunk} reports statistics on the violations found on optimized instances of the 1000 test programs. We treat violations that happen at different program lines as distinct. When a violation occurs at multiple optimization levels, we count it once in the last table row.

\begin{table}[t!]
\centering
\begin{footnotesize}
%\adjustbox{max width=0.9\linewidth}{%
\begin{tabular}{l|rrr|rrr}
%\multicolumn{1}{l}{} & \multicolumn{3}{c}{clang} & \multicolumn{3}{c}{gcc} \\
%\hline
Level & C1 & C2 & C3 & C1 & C2 & C3 \\
\hline
Og & 71 & 553 & 75 & 10 & 34 & 115 \\
O1 & - & - & - & 168 & 67 & 28 \\
O2 & 51 & 455 & 43 & 227 & 131 & 2 \\
O3 & 51 & 350 & 39 & 215 & 97 & 1  \\
Os & 73 & 471 & 52 & 233 & 141 & 1 \\
Oz & 74 & 463 & 78 & 221 & 135 & 1 \\
\hline
%total & - & - & - & 1074 & 605 & 148 \\
unique & 84 & 885 & 121 & 282 & 227 & 134
\end{tabular}
\vspace{4pt}
%}
\caption{Conjecture violations in clang (left) \& gcc (right).\label{tab:gcc-trunk}}
\vspace{-2mm}
\end{footnotesize}
\end{table}

\begin{figure}[t!]
\vspace{-2mm}
\centering
\includegraphics[width=0.745\linewidth]{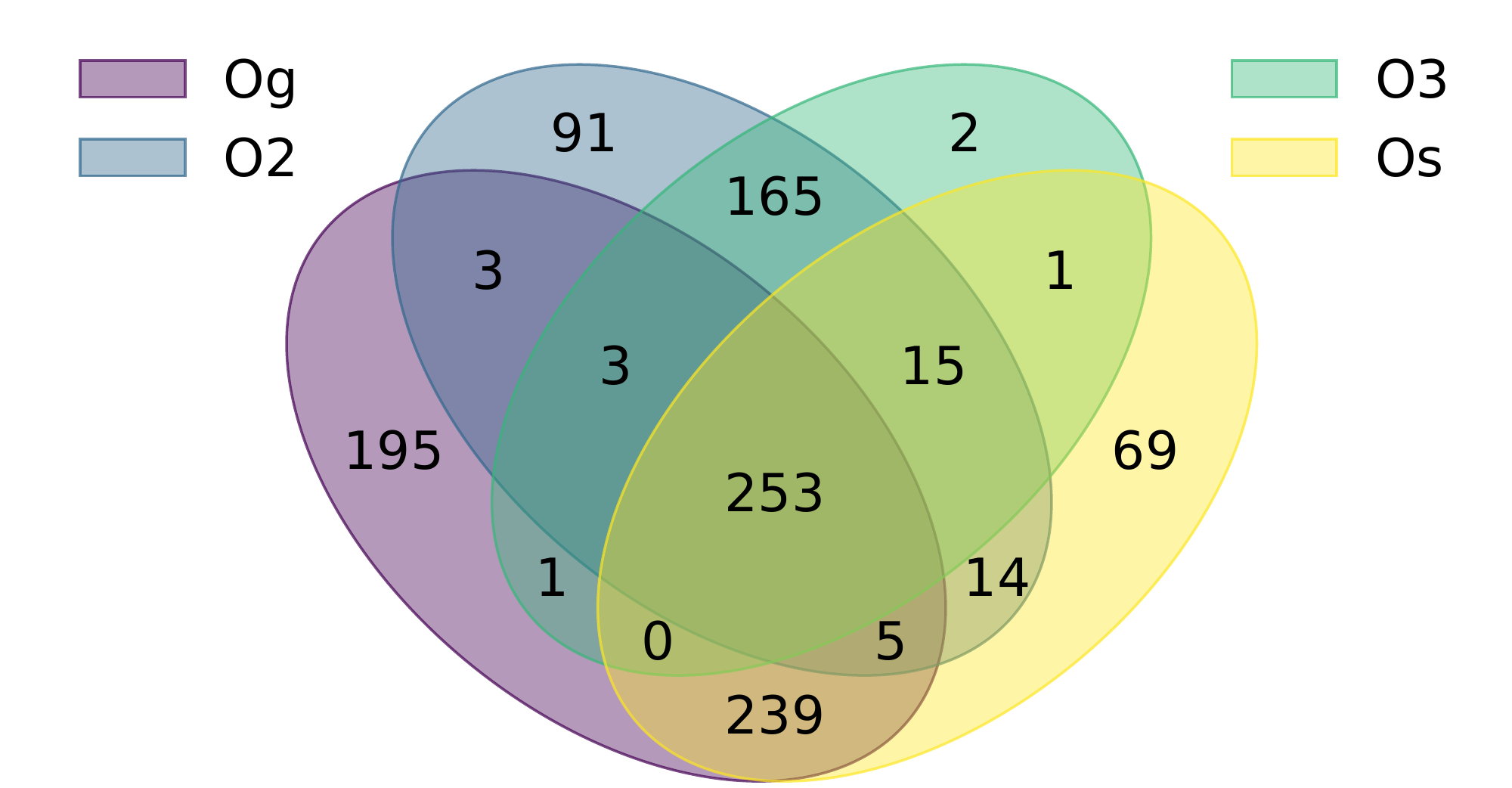}
\vspace{-2.5mm}
\caption{\label{fig:venn-clang} Unique violations for all conjectures (clang).}% from ; at different optimization levels (clang trunk, {\tt -Oz} omitted).
\end{figure}

\begin{figure}[t!]
\centering
\includegraphics[width=0.8\linewidth]{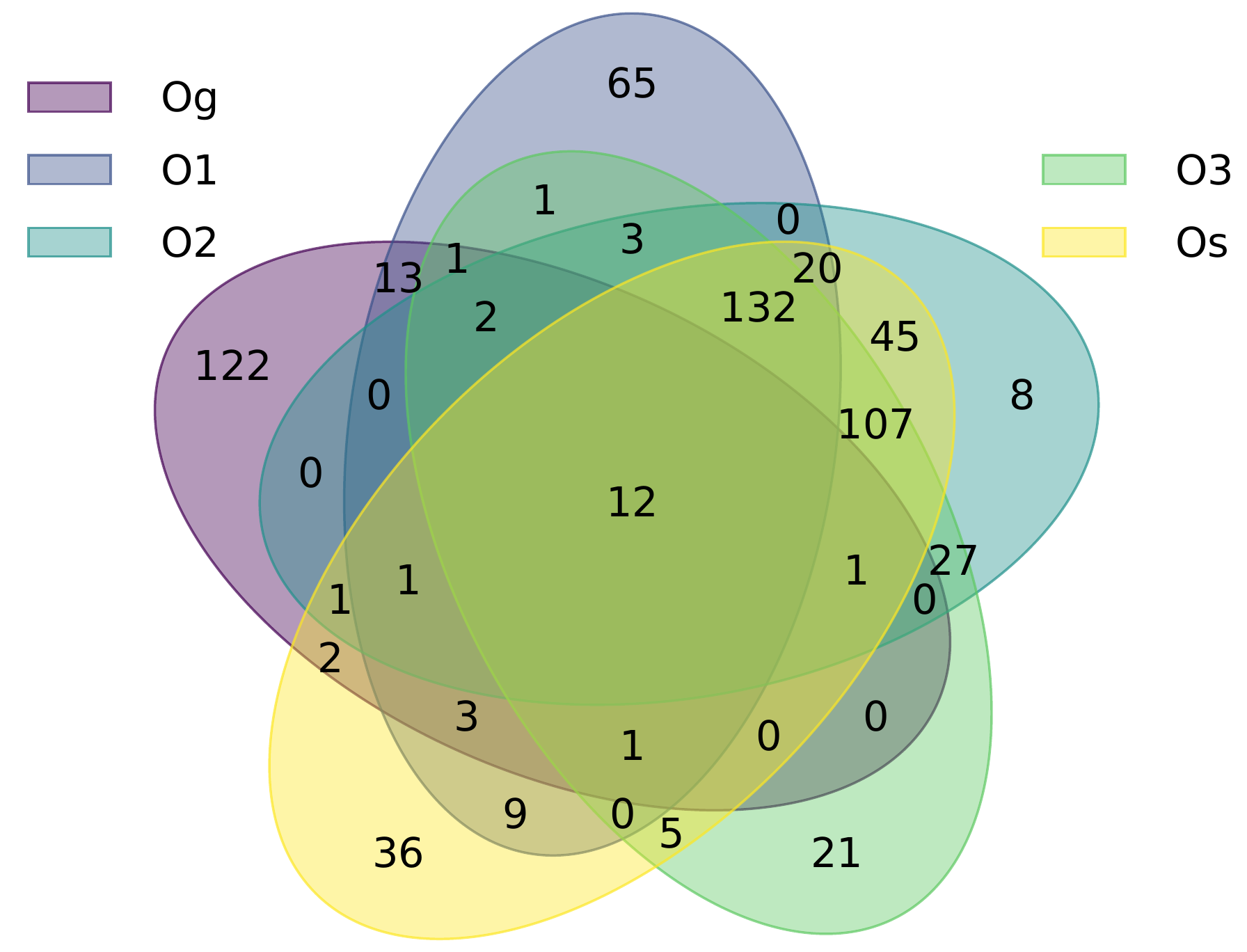}
\vspace{-2.5mm}
\caption{\label{fig:venn-gcc} Unique violations for all conjectures (gcc).}% at different optimization levels (gcc trunk, {\tt -Oz} omitted)
%\vspace{-1mm}
\end{figure}

\looseness=-1
Between compilers, a conjecture may expose very different a\-mounts of unique violations. For example, we observe way more violations for Conjecture 2 in clang than in gcc (3.9x as many) while the opposite holds for Conjecture 1 (3.36x more in gcc), while for Conjecture 3 the numbers are similar. However, violations can occur with a significantly different frequency among optimization levels.

For example, Conjecture 1 has very few violations in gcc with {\tt -Og} optimization (just 10), whereas their number increases significantly at other levels (up to 223 at {\tt -Os}). In clang, instead, the more aggressive {\tt -O2} and {\tt -O3} levels handle call arguments better than other optimization levels. We found that the optimizations that are applied at, e.g., {\tt -Og} are replaced at higher levels by more aggressive ones that, fortunately, preserve debug information better.

The clang LSR optimization bug analyzed for Conjecture 2 in Section~\ref{ss:invariant-three} impacts code generation frequently often, as loop induction variables are often used to index memory in Csmith programs (and in real-world code). We will resume its discussion in Section~\ref{ss:eval-regression}.

For Conjecture 3, we partially relate the very few violations at higher optimization levels in gcc to the drop in line ratio metric we observed in Section~\ref{se:motivation}, meaning that for lifetime inspection we can step on many fewer lines.

Finally, to put numbers in perspective, for the three conjectures we found no violations in (951, 680, 869) out of 1000 programs for clang and (846, 842, 864) for gcc.

Doing multiple optimization levels in parallel (one per core), each program was generated and tested for one conjecture in \textasciitilde30s, with no substantial variations per compiler or conjecture. Testing the 1000 programs for 3 conjectures took \textasciitilde2.5 hours per compiler.

\subsection{Heterogeneity of Violations}
\label{ss:eval-violations-reduction}
To study if our methods can stress heterogeneous components of an optimizer, we first study how the violations of the previous section map to the optimization levels where they occur.

Figure~\ref{fig:venn-clang} and Figure~\ref{fig:venn-gcc} feature Venn diagrams for clang and gcc, respectively, that plot how each unique violation reproduces at one or more optimization levels. Each counter placed at an intersection between sets represents the number of unique violations that reproduce at those optimization levels only. To keep the visualization readable, we leave out and defer the discussion of the violations that occur (also) at {\tt -Oz}. We plot violations cumulatively as we found no notable per-conjecture trends.

\begin{table}[t!]
\centering
\begin{footnotesize}
%\adjustbox{max width=0.98\linewidth}{
\setlength{\tabcolsep}{5pt}%was: 4.5
%\adjustbox{max width=0.94\linewidth}{
\begin{tabular}{|lr|lr|}
\multicolumn{2}{l|}{\textbf{gcc} (our method)} & \multicolumn{2}{l}{\textbf{clang} (opt-bisect-limit)} \\
\hline
toplevel-reorder & 57 & Inliner & 50 \\
ipa-sra & 24 & LSR & 12 \\
tree-ccp & 17 & X86 DAG->DAG InstrSel & 9 \\
tree-vrp & 15 & SimplifyCFG & 9 \\
tree-\{fre; pta\} & 11 & LoopUnroll & 1 \\
\hline\hline
toplevel-reorder (*) & 16 & LSR & 454 \\
schedule-insns2 (*) & 12 & InstCombine & 85 \\
tree-dse (*) & 8 & IPSCCP & 53 \\
tree-ch & 8 & Merge contiguous icmps & 5 \\
tree-\{loop-ivcanon; vrp\} & 7 & Canon. Freeze Instr & 5 \\
\hline\hline
ipa-pure-const & 20 & InstCombine & 18 \\
tree-ccp & 11 & X86 DAG->DAG InstrSel & 17 \\
tree-dce & 9 & SROA & 15 \\
tree-fre & 4 & Machine InstrScheduler & 11 \\
ipa-refer.-addressable (*) & 3 & PostOrderFunctionAttrs & 11  \\
\hline
\end{tabular}
%}
\vspace{4pt}
\end{footnotesize}
\caption{\label{tab:topk}Triaged optimizations (top-5 only). Conjectures 1, 2, and 3 are ordered vertically. (*) means after inlining.}
\end{table}

For clang, out of 1056 unique violations, about a fourth (253) occur at all optimization levels, 195 only at {\tt -Og} and 239 only at {\tt -Og} and {\tt -Os}. While many transformations are shared, others activate instead only for some levels.

For gcc, out of 638 unique violations, very few (12) occur at all levels, 122 only at {\tt -Og}, 65 only at {\tt -O1}, but even more interestingly 132 at all levels except {\tt -Og} and {\tt 107} at all levels but {\tt -O1} and {\tt -Og}. These trends are almost anti-symmetric with what we observed in clang and may be explained in different ways, including substantial differences\footnotemark{} in what optimizations each level applies.

For violations occurring only at {\tt -Oz}, which optimizes for size more aggressively than {\tt -Os}, we found (6, 6, 22) of them in clang and (0, 5, 0) in gcc for the three conjectures.

\begin{table}[t!]
\centering
\begin{footnotesize}
\begin{tabular}{|c|c|c|c|c|}
    \toprule
    Tracker ID & System & Bug status & Conjecture & DWARF analysis \\
    \midrule
    49546 & clang & Confirmed & C1 & Missing DIE \\
    49580 & clang & Confirmed & C1 & Missing DIE \\
    49769 & clang & Confirmed & C1 & Hollow DIE  \\
    49973 & clang & Confirmed & C1 & Hollow DIE  \\
    49975 & clang & Confirmed & C1 & Hollow DIE  \\
    51780 & clang  & Confirmed & C1 & Missing DIE \\
    55101 & clang & Unconfirmed & C1 & Hollow DIE  \\
    55115 & clang & Confirmed & C1 & Missing DIE \\
    55123 & clang & Unconfirmed & C1 & Hollow DIE  \\
    53855a & clang & Fixed by {\tt trunk*} & C2 & Hollow DIE  \\
    53855b & clang & Confirmed & C2 & Hollow DIE  \\
    54611 & clang & Unconfirmed & C2 & Incomplete DIE \\
    54757 & clang & Unconfirmed & C2 & Hollow DIE  \\
    54763 & clang & Unconfirmed & C2 & Incomplete DIE  \\
    50286 & clang & Confirmed & C3 & Incomplete DIE \\
    54796 & clang & Confirmed & C3 & Incomplete DIE \\
    \midrule
    104549 & gcc & Unconfirmed & C1 & Incorrect DIE \\
    105007 & gcc & Confirmed & C1 & Hollow DIE  \\
    105158 & gcc & Fixed & C1 & Hollow DIE  \\
    105176 & gcc & Unconfirmed & C1 & Incomplete DIE \\
    105179 & gcc & Unconfirmed & C1 & Incomplete DIE \\
    105239 & gcc & Unconfirmed & C1 & Incomplete DIE \\
    105248 & gcc & Confirmed & C1 & Hollow DIE  \\
    105261 & gcc & Confirmed & C1 & Hollow DIE  \\
    104891 & gcc & Unconfirmed & C2 & Incomplete DIE \\
    105036 & gcc & Unconfirmed & C2 & Incorrect DIE \\
    105108 & gcc & Confirmed & C2 & Hollow DIE  \\
    105145 & gcc & Confirmed & C2 & Hollow DIE  \\
    105161 & gcc & Confirmed & C2 & Hollow DIE  \\
    105249 & gcc & Unconfirmed & C2 & Incorrect DIE \\
    104938 & gcc & Confirmed & C3 & Incomplete DIE \\
    105124 & gcc & Confirmed & C3 & Incomplete DIE \\
    105159 & gcc & Unconfirmed & C3 & Hollow DIE  \\
    105194 & gcc & Fixed & C3 & Incomplete DIE \\
    105389 & gcc & Unconfirmed & C3 & Incomplete DIE  \\
    \midrule
    28987 & gdb & Confirmed & C1 & - \\
    29060 & gdb & Confirmed & C1 & - \\
    \midrule
    50076 & lldb & Confirmed & C1 & - \\ % 50076 is the one that Jeremy filed after, 49973 ours in clang
    \bottomrule
\end{tabular}
\end{footnotesize}
\vspace{2pt}
\caption{Reported issues and their current status.\label{tab:reports}}
\vspace{-2pt}
\end{table}

\footnotetext{For our triaging method we identified 81, 94, 138, 151, and 131 boolean flags that impact optimization in {\tt O\{g,1,2,3,s\}}, respectively.}

Moving on, we grouped all the found violations using the techniques presented in Section~\ref{ss:multi-scz}. Table~\ref{tab:topk} shows the five transformations that we identified as most frequently behind the violations. We see that transformations involving instruction scheduling (as anticipated already in~\cite{AdlTabatabai-PLDI93}), strength reduction, and loops recurrently violate multiple conjectures. In gcc, we note that tree-based optimizations done in its GIMPLE framework also occur frequently. We found a total of 31 distinct clang passes (8, 20, and 22 for the three conjectures) and 271 unique boolean flags and combinations thereof for gcc (68, 178, 49) in all tests\footnote{Besides a possible higher sparsity of implementation defects across transformations in gcc, we explain this difference also with the search method: the native incremental bisection of LLVM halts on the first pass causing a regression whereas we try all the search space in gcc, exposing cases where distinct parts can cause the same final effect.}.

Identifying the culprit transformation took on average 20 minutes per program with gcc and 4 with clang (with one core). Test minimization in C-Reduce took \textasciitilde1 hour for most violations on both clang and gcc (with 20 cores).

\subsection{Notable Traits of Found Issues}
\label{ss:eval-bugs}

Table~\ref{tab:reports} shows the 38 issues we reported so far to the developers of the LLVM (clang+lldb) and GNU (gcc+gdb) ecosystems, resulting in 11 confirmed bugs for clang (8 passes), 10 for gcc (7 transformations), 1 for lldb, and 2 for gdb, for a total of 24. Other reports (5 for clang and 9 for gcc) await analysis or determinations. The conjectures revealed, respectively, 20 (14 confirmed), 11 (5), and 7 (5) issues.

From a debugging experience perspective, the violations are evenly split between variables marked as optimized out and variables not visible at all. If we then look at the nature of the affected variables, about half of the violations involve variables holding a constant value, that is, coming from an assignment with a literal or a constant-folded expression (as a result of one or more optimizations). For such a variable, the optimizer generally avoids storage allocation in the code but can emit a DWARF {\tt DW\_AT\_const\_value} attribute in the DIE of the variable to make it available during debugging. The remaining violations are evenly split between variables kept in registers and/or memory and variables for which the optimizer could avoid storage allocation as they host different constant values at different location ranges\footnotemark{}. Missing information in all said remaining violations could have been encoded using the {\tt DW\_AT\_location} attribute in the associated variable DIEs.

Analyzing the Debug Information Entry (DIE) of a variable can be helpful to collect further evidence on how the compiler mishandled it. We can divide the 35 compiler-related issues in four categories:
\begin{itemize}
    \item \textbf{Missing DIE} (4 issues): the information the debugger accesses at the program point contains no DIE for the variable;
    \item \textbf{Hollow DIE} (16 issues): the optimizer is aware of the variable but its DIE shows no location or constant-value information (i.e., neither above-discussed attribute is present);
    \item \textbf{Incomplete DIE} (12 issues): the location definition present in the DIE of the variable does not cover all the instructions related to source-level lines where the variable is visible.
    \item \textbf{Incorrect DIE} (3 issues): the optimizer tracked the variable in full but the debugger cannot display its value due to incorrect DIE information for the program point(s) involved.
\end{itemize}

\paragraph{Case Studies} We discuss below exemplary bugs for each category, as well as also one of the reported issues (3) for the debugger tools. Appendix~\ref{se:appendix} provides brief analyses of all the 38 issues we found.

\smallskip
{\bf Missing DIE:} Bug 49546 in clang involves an induction variable {\tt j} from a {\tt for (j=0; j<1; j++)} loop passed to an opaque function called by the loop. After loop rotation, the compiler realizes that only one iteration will take place: as a clang developer observed~\cite{clang-49546}, it should be possible for the optimizer to mark one region of the function to show {\tt j=0} and another {\tt j=1} when debugging. However, debug information is eventually lost for both program points related to the assignment (due to loop optimization effects for {\tt j=1} and for a bug in the common {\tt SimplifyCFG} pass for {\tt j=0}) and so is the DIE.

{\bf Hollow DIE:} Bug 105108 in gcc involves assigning a variable with an expression that, among its constituents, includes the return value a function doing just {\tt return 0}. At {\tt -O1/-Og}, the optimizer can constant-fold the expression thanks to conditional constant propagation and value range propagation, while at {\tt -O2/-O3} inlining makes the constant-folding trivial. All these optimization levels eventually produce the same code for the program, but only with {\tt -O1/-Og} the DIE misses the {\tt DW\_AT\_const\_value} attribute~\cite{bug-gcc-105108}.

{\bf Incomplete DIE:} Bug 105179 in gcc involves a variable declared outside a loop, where the latter in turn assigns and uses the former twice as call argument: once to a function in the same code module and once to an opaque one. When compiling at {\tt -Og}, the variable is displayed with its value only at the first function call, whereas it displays at both calls with other optimization levels. We found that the {\tt -fcprop-registers} transformation (a copy-propagation pass to reduce scheduling dependencies) leads to a range for the variable that does not cover the address of the call, despite the optimizer is aware of where the variable is stored~\cite{bug-gcc-105179}.

\footnotetext{This can occur often with heavy-duty loop transformations enabled by unrolling.}

{\bf Incorrect DIE:} Bug 105249 in gcc involves a zero-initialized induction variable that, when compiling with {\tt -Os}, is not visible in the body of a loop {\tt for (;i<2;i++) a=b[i];} that manipulates storage {\tt volatile int a; int b[2];} from global memory. We spotted wrong location information from when the unrolled loop body undergoes instruction scheduling, which erroneously associates the instructions with the DIE of an inlined function called right after the loop. Even if the DIE for {\tt i} correctly keeps track of its value using the DWARF expression stack, the debugger cannot display it since {\tt i} is not part of the frame of the inlined function~\cite{bug-gcc-105249}.

{\bf Debugger tools:} Bug 50076 in lldb involves a variable used as call argument to an opaque function; the call takes place in a function that the clang compiler inlines into {\tt main()}. As we found in our initial violation investigation (Section~\ref{ss:checking-conjectures}), the issue does not occur when analyzing the executable with gdb. A developer followed up on our test code~\cite{lldb-50076}, noting that lldb may not be able to show variables that appear only in the abstract origins of {\tt DW\_TAG\_inlined\_subroutine}.

\paragraph{Discussion} We find that the issues described above and in the past sections suggest that debug-related issues are scattered among the components of an optimizer compiler. Systematic testing approaches like the one we propose can be helpful for developers, relieving them from the burden of finding these violations manually.

Some of our reports sparked interesting discussions.
In the 105108 Hollow DIE case, the developers of gcc noted that its current design inevitably loses track of the call to the incriminated function (as a consequence of detecting it as pure---i.e., side effect-free---and thus deleting it) unless inlined. They then discussed a potential a DWARF 6 addition so that the affected function may be expressed in DWARF bytecode and invoked to recover the result~\cite{bug-gcc-105108}. Report 105145 for gcc~\cite{bug-gcc-105145} brought to the surface a design limitation in retaining debug information for memory contents involving address-taken local variables that eventually get stored in registers. Similar gaps were acknowledged for clang too (for example, in issues 51780~\cite{clang-51780} and 55115~\cite{clang-55115}). Sometimes, our programs exercised patterns that existing provisions did not handle properly (e.g., clang issue 53855~\cite{bug-clang-53855}, gcc issue 105161~\cite{bug-gcc-105161}, gdb issue 28987~\cite{gdb-28987}).

\begin{figure*}[ht!]
\centering
\includegraphics[width=\linewidth]{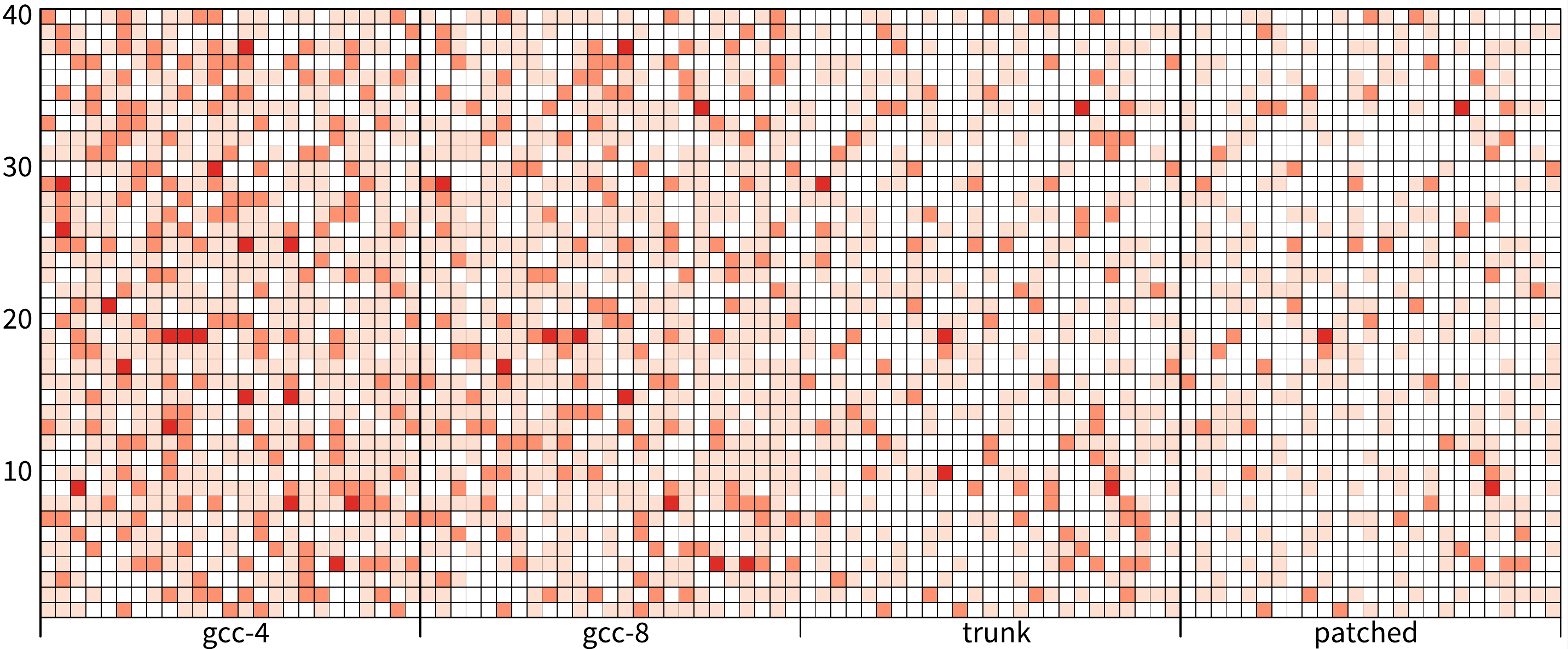}%
\vspace{-2pt}
\caption{\label{fig:gcc-regression} Number of conjectures violated by a test program on different gcc versions. The 1000 programs are arranged as 25 per row. The color code is as follows: \fcolorbox{black}[HTML]{FFFFFF}{\rule{0pt}{3pt}} 0 conjectures, \fcolorbox{black}[HTML]{FEE0D2}{\rule{0pt}{3pt}} 1 conjecture, \fcolorbox{black}[HTML]{FC9272}{\rule{0pt}{3pt}} 2 conjectures, \fcolorbox{black}[HTML]{DE2D26}{\rule{0pt}{3pt}} 3 conjectures.}
\vspace{2pt}
\end{figure*}

Across all reported compiler-related issues, we noted higher engagement in bugs occurring only at one optimization level, especially {\tt -Og}. For the 14 unconfirmed bugs, at the time of writing none has been rejected and 5 have received a preliminary answer; some have seen only the addition in the bug tracker of a tag for the involved component and/or have been referenced in other reports. Among all reported issues, we noted reaction times varying from same-day confirmation to a 2-week-or-longer wait.
Besides possibly different levels of interest among developers, their workload may have been a factor too, as it happened with one acknowledged hiatus case. We found instead no evidence relating the (current) lack of a follow-up on a report to the technicalities behind the issue described.
As for the already patched bugs, we noted that they involved extending or changing how a specific transformation moves debug metadata during basic block manipulations. On the other hand, outstanding bugs involving infrastructure limitations or requiring synchronized changes in multiple components may possibly just need longer to be addressed.

Ultimately, we found the developers reaction promising and received rather valuable feedback back. Future practical research may possibly focus on optimizing the reporting activity by providing additional context for speeding up bug analysis: for example, by tracking what parts of the affected transformations are exercised by the test program (e.g., leveraging the infrastructure released by the authors of~\cite{Karine22}) or by trying to generate multiple variants of it.

\subsection{Regression Study}
\label{ss:eval-regression}
For bug 105158 that we reported for gcc~\cite{bug-gcc-105158}, a developer wrote a patch that we use in the following to discuss the potential benefits of our testing. Albeit one cannot generalize from a single experience, we find the results we are about to present quite suggestive. Table~\ref{tab:regression} shows how the number of unique violations for the three conjectures are affected by a single change.

The {\tt cleanup\_tree\_cfg} helper is shared by many transformations in gcc. A violation of the first conjecture exposed the flaw. After the patch, the violations for it drop by 63.5\% (from 282 to 131), followed by some improvements for the two other conjectures (5.2\% and 5.9\%). Being a helper, the unique (combinations of) transformations behind the fixed violations were 68. For the availability-of-variables metric from the study of Section~\ref{se:motivation}, the value moved up from 0.8562 to 0.8633 for {\tt -O1}, bridging almost half of the gap with {\tt -Og} (0.8758).

While all our clang bugs await fixing, we study a concurrent partial fix that its developers wrote for LSR (Section~\ref{ss:invariant-three}). We use a later commit 796b84d, dubbed trunk* in Table~\ref{tab:regression} and focus only the violations from Section~\ref{ss:eval-violations-trunk} that we traced back to LSR. We observe an 80.4\% reduction for them (from 454 to 89), which hopefully will be further improved after our feedback (Section~\ref{ss:invariant-three}).

%\looseness=-1
Table~\ref{tab:regression} also shows how the violations significantly vary for both compilers when taking versions far apart in time. Typically, they decrease for all conjectures, confirming our beliefs from Section~\ref{se:motivation} on the ongoing improvements in compilation systems. Incidentally, we note some regressions for Conjecture 3 on trunk*. As a teaser of how our proposal may help developers track regressions, Figure~\ref{fig:gcc-regression} shows the conjectures violated on the fixed 1000 programs in several gcc versions (we omit clang for brevity).

\begin{table}
\centering
\begin{footnotesize}
\setlength\tabcolsep{4pt}
\begin{tabular}{l|rrrr|rrrr}
& gcc4 & gcc8 & trunk & patched & clang5 & clang9 & trunk & trunk* \\
\hline
C1 & 277 & 180 & 282 & 131 & 30 & 10 & 84 & 70 \\
C2 & 1259 & 962 & 227 & 215  & 1498 & 1297 & 885 & 518 \\
C3 & 168 & 134 & 134 & 126  & 329 & 196 & 121 & 124 \\
\hline
\end{tabular}
\end{footnotesize}
\vspace{2pt}
\caption{\label{tab:regression} Number of violations across compiler versions.}
\end{table}

% !TEX root = main.tex
\section{Related Works}
\label{se:related}

\paragraph{Testing of Debug Information.}
The work of Li et al.~\cite{Davide-PLDI20} is the first attempt to systematically test the correctness of debug information, in particular for variable values. The method generates test programs for which one can step on specific lines and validate the values of selected variables---crafted to be unoptimizable---by using an unoptimized executable instance as oracle.

Di Luna et al.~\cite{Giuseppe-ASPLOS21} do away with ad-hoc program generation and identify four general properties (``invariants'') involving different aspects of the information presented to users (e.g., spurious frames, out-of-scope variables).
We acknowledge implementations similarities with their work, as using debuggers to check behaviors and compiler fuzzers to generate test programs are choices that turned out to be effective for us too. The conjectures we propose here are also similar in the spirit to their invariants. Scientifically, instead, the works fundamentally differ in the object and in the technique of the analysis. For the former, \cite{Giuseppe-ASPLOS21} validates existing debugging information whereas we study when absence of information is attributable to bugs. For the latter, \cite{Giuseppe-ASPLOS21} can rely on the ground truth given by {\tt -O0}, which simplifies the theoretical grounds and makes validation straightforward; the problem we study comes with no baseline or oracle and we are the first to spot and present it as such.

More recently, Artuso et al~\cite{Fiorella-Acc22} propose neural-based techniques to identify discrepancies in the mapping between source-level locations and assembly code in debug information. Again, validation comes here from the straightforward ground truth given by {\tt -O0}.

\paragraph{Debugging Optimized Code.} Researchers have been aware of the conflict between the (inevitable) effects of optimizations and the (desirable) ability to debug a program in terms of its source representation for a long time now (e.g.,~\cite{Hennessy-TOPLAS82, Brooks-PLDI92, Copperman-TOPLAS94}). Some studies have analyzed classes of optimizations that make variables \textit{endangered} (i.e., their run-time value may be inconsistent with the source-level value expected at a breakpoint~\cite{AdlTabatabai-PLDI96}) and proposed techniques (e.g.,~\cite{LeChun-PLDI99, Jaramillo-SAS00, Daniele-PLDI18}) to reconstruct expected values under specific optimizations. While the restrictive assumptions they make limit their applicability to mainstream optimizers, in our scenario we may consider them as a means to recover debug information losses from unavoidable effects of optimization.

A promising direction could be to turn some of the ideas on their head to identify, for example, classes of ``non-endangered'' variables that one would expect to be available. To some extent, Conjecture 2 already embodies this flavor, as it studies constituents representing constants or types of unoptimizable values.

% !TEX root = main.tex
\section{Limitations and Future Works}
\label{se:limitations}
%\looseness=-1
We acknowledge the following limitations and threats to validity.

We make use of dynamic analysis in debuggers to identify violations: if a program line does not execute or if the debugger mishandles a case, our results are affected. To mitigate this risk, we use a compiler fuzzer to generate input-independent test programs and resort to multiple debuggers for validating violations (Section~\ref{ss:checking-conjectures}). We may also miss violations due to conservative choices in the checking logic: e.g., by failing to identify a constituent variable that Conjecture 2 expects to be available. Conservative provisions mitigate the risk of false positives, whereas more complex analyses on the source code or the assembly generated for it could allow us to relax some and hopefully expose more defects. Finally, while the conjectures proposed and analyzed in this paper are designed to expose only bug-induced violations, their empirically derived nature inevitably makes them only ``likely correct''.

%\looseness=-1
The presented conjecture examples do not cover the possibility of stepping on specific lines. In general, two compilers may decide differently in whether to generate code for a line (Section~\ref{se:motivation}). However, sometimes, the visibility of a line may imply that code for a related line exists. We played with control dependencies~\cite{Cytron-TPLS91} by searching for any of the lines that a line is control-dependent on: while early tests were inconclusive, future work could pick up this idea. Nonetheless, by making ``more'' variables available as we seek, the benefits not only are immediate for user debugging experience, but also transfer to new lines that may be recovered in other ways.
 
We would also like to test multiple architectures. We filed one issue~\cite{report-llvm-arch64} where the violation on x86\_64 did not reproduce in clang on aarch64 due to differences in the backends. Our pipeline can be readily adapted for systematic multi-architecture testing by using, for example, the fast user-mode emulation of QEMU for testing executables for different architectures at once~\cite{DTRAP22}.

%\textcolor{brown}{To Robert}\textcolor{cyan}{, for the bagels} \textcolor{pink}{and explaining CMYK} \textcolor{green}{and color spaces.}
\begin{acks}
We are indebted to Davide Italiano for the rich discussions about identifying the debug information completeness problem and the technicalities backing Conjecture 1 and 3. We thank our anonymous reviewers and particularly our shepherd for the rich feedback and the guidance we received for improving the manuscript.

This work has been partially supported by the IoT-STYLE project RG12117A7CE68848 and by the SAFE (Self-attentive function embeddings for embedded systems) project.
\end{acks}


\bibliographystyle{ACM-Reference-Format}
\bibliography{references}


\begin{thebibliography}{42}

%%% the Web address.

\ifx \showCODEN    \undefined \def \showCODEN     #1{\unskip}     \fi
\ifx \showDOI      \undefined \def \showDOI       #1{#1}\fi
\ifx \showISBNx    \undefined \def \showISBNx     #1{\unskip}     \fi
\ifx \showISBNxiii \undefined \def \showISBNxiii  #1{\unskip}     \fi
\ifx \showISSN     \undefined \def \showISSN      #1{\unskip}     \fi
\ifx \showLCCN     \undefined \def \showLCCN      #1{\unskip}     \fi
\ifx \shownote     \undefined \def \shownote      #1{#1}          \fi
\ifx \showarticletitle \undefined \def \showarticletitle #1{#1}   \fi
\ifx \showURL      \undefined \def \showURL       {\relax}        \fi
% invisible to TeX
\providecommand\bibfield[2]{#2}
\providecommand\bibinfo[2]{#2}
\providecommand\natexlab[1]{#1}
\providecommand\showeprint[2][]{arXiv:#2}

\bibitem[Adl-Tabatabai and Gross(1993)]%
        {AdlTabatabai-PLDI93}
\bibfield{author}{\bibinfo{person}{Ali-Reza Adl-Tabatabai} {and}
  \bibinfo{person}{Thomas Gross}.} \bibinfo{year}{1993}\natexlab{}.
\newblock \showarticletitle{Detection and Recovery of Endangered Variables
  Caused by Instruction Scheduling}. In \bibinfo{booktitle}{\emph{Proc. of the
  ACM SIGPLAN 1993 Conference on Programming Language Design and
  Implementation}} (Albuquerque, New Mexico, USA) \emph{(\bibinfo{series}{PLDI
  '93})}. \bibinfo{publisher}{Association for Computing Machinery},
  \bibinfo{pages}{13–25}.
\newblock
\showISBNx{0897915984}
\urldef\tempurl%
\url{https://doi.org/10.1145/155090.155092}
\showDOI{\tempurl}


\bibitem[Adl-Tabatabai and Gross(1996)]%
        {AdlTabatabai-PLDI96}
\bibfield{author}{\bibinfo{person}{Ali-Reza Adl-Tabatabai} {and}
  \bibinfo{person}{Thomas Gross}.} \bibinfo{year}{1996}\natexlab{}.
\newblock \showarticletitle{Source-Level Debugging of Scalar Optimized Code}.
  In \bibinfo{booktitle}{\emph{Proc. of the ACM SIGPLAN 1996 Conference on
  Programming Language Design and Implementation}} (Philadelphia, Pennsylvania,
  USA) \emph{(\bibinfo{series}{PLDI '96})}. \bibinfo{publisher}{Association for
  Computing Machinery}, \bibinfo{pages}{33–43}.
\newblock
\showISBNx{0897917952}
\urldef\tempurl%
\url{https://doi.org/10.1145/231379.231388}
\showDOI{\tempurl}


\bibitem[Artuso et~al\mbox{.}(2022)]%
        {Fiorella-Acc22}
\bibfield{author}{\bibinfo{person}{Fiorella Artuso},
  \bibinfo{person}{Giuseppe~Antonio Di~Luna}, {and} \bibinfo{person}{Leonardo
  Querzoni}.} \bibinfo{year}{2022}\natexlab{}.
\newblock \showarticletitle{Debugging Debug Information With Neural Networks}.
\newblock \bibinfo{journal}{\emph{IEEE Access}}  \bibinfo{volume}{10}
  (\bibinfo{year}{2022}), \bibinfo{pages}{54136--54148}.
\newblock
\urldef\tempurl%
\url{https://doi.org/10.1109/ACCESS.2022.3176617}
\showDOI{\tempurl}


\bibitem[Brooks et~al\mbox{.}(1992)]%
        {Brooks-PLDI92}
\bibfield{author}{\bibinfo{person}{Gary Brooks}, \bibinfo{person}{Gilbert~J.
  Hansen}, {and} \bibinfo{person}{Steve Simmons}.}
  \bibinfo{year}{1992}\natexlab{}.
\newblock \showarticletitle{A New Approach to Debugging Optimized Code}. In
  \bibinfo{booktitle}{\emph{Proc. of the ACM SIGPLAN 1992 Conference on
  Programming Language Design and Implementation}} (San Francisco, California,
  USA) \emph{(\bibinfo{series}{PLDI '92})}. \bibinfo{publisher}{Association for
  Computing Machinery}, \bibinfo{pages}{1–11}.
\newblock
\showISBNx{0897914759}
\urldef\tempurl%
\url{https://doi.org/10.1145/143095.143108}
\showDOI{\tempurl}


\bibitem[Chen et~al\mbox{.}(2016)]%
        {CompBugs-ICSE16}
\bibfield{author}{\bibinfo{person}{Junjie Chen}, \bibinfo{person}{Wenxiang Hu},
  \bibinfo{person}{Dan Hao}, \bibinfo{person}{Yingfei Xiong},
  \bibinfo{person}{Hongyu Zhang}, \bibinfo{person}{Lu Zhang}, {and}
  \bibinfo{person}{Bing Xie}.} \bibinfo{year}{2016}\natexlab{}.
\newblock \showarticletitle{An Empirical Comparison of Compiler Testing
  Techniques}. In \bibinfo{booktitle}{\emph{Proc. of the 38th International
  Conference on Software Engineering}} (Austin, Texas)
  \emph{(\bibinfo{series}{ICSE '16})}. \bibinfo{publisher}{Association for
  Computing Machinery}, \bibinfo{pages}{180–190}.
\newblock
\showISBNx{9781450339001}
\urldef\tempurl%
\url{https://doi.org/10.1145/2884781.2884878}
\showDOI{\tempurl}


\bibitem[Chen et~al\mbox{.}(2013)]%
        {CompBugs-PLDI13}
\bibfield{author}{\bibinfo{person}{Yang Chen}, \bibinfo{person}{Alex Groce},
  \bibinfo{person}{Chaoqiang Zhang}, \bibinfo{person}{Weng-Keen Wong},
  \bibinfo{person}{Xiaoli Fern}, \bibinfo{person}{Eric Eide}, {and}
  \bibinfo{person}{John Regehr}.} \bibinfo{year}{2013}\natexlab{}.
\newblock \showarticletitle{Taming Compiler Fuzzers}. In
  \bibinfo{booktitle}{\emph{Proc. of the 34th ACM SIGPLAN Conference on
  Programming Language Design and Implementation}} (Seattle, Washington, USA)
  \emph{(\bibinfo{series}{PLDI '13})}. \bibinfo{publisher}{Association for
  Computing Machinery}, \bibinfo{pages}{197–208}.
\newblock
\showISBNx{9781450320146}
\urldef\tempurl%
\url{https://doi.org/10.1145/2491956.2462173}
\showDOI{\tempurl}


\bibitem[Copperman(1994)]%
        {Copperman-TOPLAS94}
\bibfield{author}{\bibinfo{person}{Max Copperman}.}
  \bibinfo{year}{1994}\natexlab{}.
\newblock \showarticletitle{Debugging Optimized Code without Being Misled}.
\newblock \bibinfo{journal}{\emph{ACM Trans. Program. Lang. Syst.}}
  \bibinfo{volume}{16}, \bibinfo{number}{3} (\bibinfo{date}{may}
  \bibinfo{year}{1994}), \bibinfo{pages}{387–427}.
\newblock
\showISSN{0164-0925}
\urldef\tempurl%
\url{https://doi.org/10.1145/177492.177517}
\showDOI{\tempurl}


\bibitem[Cummins et~al\mbox{.}(2018)]%
        {CompBugs-ISSTA18}
\bibfield{author}{\bibinfo{person}{Chris Cummins}, \bibinfo{person}{Pavlos
  Petoumenos}, \bibinfo{person}{Alastair Murray}, {and} \bibinfo{person}{Hugh
  Leather}.} \bibinfo{year}{2018}\natexlab{}.
\newblock \showarticletitle{Compiler Fuzzing through Deep Learning}. In
  \bibinfo{booktitle}{\emph{Proc. of the 27th ACM SIGSOFT International
  Symposium on Software Testing and Analysis}} (Amsterdam, Netherlands)
  \emph{(\bibinfo{series}{ISSTA 2018})}. \bibinfo{publisher}{Association for
  Computing Machinery}, \bibinfo{pages}{95–105}.
\newblock
\showISBNx{9781450356992}
\urldef\tempurl%
\url{https://doi.org/10.1145/3213846.3213848}
\showDOI{\tempurl}


\bibitem[Cytron et~al\mbox{.}(1991)]%
        {Cytron-TPLS91}
\bibfield{author}{\bibinfo{person}{Ron Cytron}, \bibinfo{person}{Jeanne
  Ferrante}, \bibinfo{person}{Barry~K. Rosen}, \bibinfo{person}{Mark~N.
  Wegman}, {and} \bibinfo{person}{F.~Kenneth Zadeck}.}
  \bibinfo{year}{1991}\natexlab{}.
\newblock \showarticletitle{Efficiently Computing Static Single Assignment Form
  and the Control Dependence Graph}.
\newblock \bibinfo{journal}{\emph{ACM Trans. Program. Lang. Syst.}}
  \bibinfo{volume}{13}, \bibinfo{number}{4} (\bibinfo{date}{oct}
  \bibinfo{year}{1991}), \bibinfo{pages}{451–490}.
\newblock
\showISSN{0164-0925}
\urldef\tempurl%
\url{https://doi.org/10.1145/115372.115320}
\showDOI{\tempurl}


\bibitem[D'Elia and Demetrescu(2018)]%
        {Daniele-PLDI18}
\bibfield{author}{\bibinfo{person}{Daniele~Cono D'Elia} {and}
  \bibinfo{person}{Camil Demetrescu}.} \bibinfo{year}{2018}\natexlab{}.
\newblock \showarticletitle{On-Stack Replacement, Distilled}. In
  \bibinfo{booktitle}{\emph{Proc. of the 39th ACM SIGPLAN Conference on
  Programming Language Design and Implementation}} (Philadelphia, PA, USA)
  \emph{(\bibinfo{series}{PLDI 2018})}. \bibinfo{publisher}{Association for
  Computing Machinery}, \bibinfo{pages}{166–180}.
\newblock
\showISBNx{9781450356985}
\urldef\tempurl%
\url{https://doi.org/10.1145/3192366.3192396}
\showDOI{\tempurl}


\bibitem[D'Elia et~al\mbox{.}(2022)]%
        {DTRAP22}
\bibfield{author}{\bibinfo{person}{Daniele~Cono D'Elia},
  \bibinfo{person}{Lorenzo Invidia}, \bibinfo{person}{Federico Palmaro}, {and}
  \bibinfo{person}{Leonardo Querzoni}.} \bibinfo{year}{2022}\natexlab{}.
\newblock \showarticletitle{Evaluating Dynamic Binary Instrumentation Systems
  for Conspicuous Features and Artifacts}.
\newblock \bibinfo{journal}{\emph{Digital Threats}} \bibinfo{volume}{3},
  \bibinfo{number}{2}, Article \bibinfo{articleno}{10} (\bibinfo{date}{feb}
  \bibinfo{year}{2022}), \bibinfo{numpages}{13}~pages.
\newblock
\showISSN{2692-1626}
\urldef\tempurl%
\url{https://doi.org/10.1145/3478520}
\showDOI{\tempurl}


\bibitem[Di~Luna et~al\mbox{.}(2021)]%
        {Giuseppe-ASPLOS21}
\bibfield{author}{\bibinfo{person}{Giuseppe~Antonio Di~Luna},
  \bibinfo{person}{Davide Italiano}, \bibinfo{person}{Luca Massarelli},
  \bibinfo{person}{Sebastian \"{O}sterlund}, \bibinfo{person}{Cristiano
  Giuffrida}, {and} \bibinfo{person}{Leonardo Querzoni}.}
  \bibinfo{year}{2021}\natexlab{}.
\newblock \showarticletitle{Who’s Debugging the Debuggers? {Exposing} Debug
  Information Bugs in Optimized Binaries}. In \bibinfo{booktitle}{\emph{Proc.
  of the 26th ACM International Conference on Architectural Support for
  Programming Languages and Operating Systems}} (Virtual, USA)
  \emph{(\bibinfo{series}{ASPLOS 2021})}. \bibinfo{publisher}{Association for
  Computing Machinery}, \bibinfo{pages}{1034–1045}.
\newblock
\showISBNx{9781450383172}
\urldef\tempurl%
\url{https://doi.org/10.1145/3445814.3446695}
\showDOI{\tempurl}


\bibitem[D'Silva et~al\mbox{.}(2015)]%
        {DSilva-SP15}
\bibfield{author}{\bibinfo{person}{Vijay D'Silva}, \bibinfo{person}{Mathias
  Payer}, {and} \bibinfo{person}{Dawn Song}.} \bibinfo{year}{2015}\natexlab{}.
\newblock \showarticletitle{The Correctness-Security Gap in Compiler
  Optimization}. In \bibinfo{booktitle}{\emph{2015 IEEE Security and Privacy
  Workshops}}. \bibinfo{pages}{73--87}.
\newblock
\urldef\tempurl%
\url{https://doi.org/10.1109/SPW.2015.33}
\showDOI{\tempurl}


\bibitem[Ernst et~al\mbox{.}(1999)]%
        {Ernst-ICSE99}
\bibfield{author}{\bibinfo{person}{Michael~D. Ernst}, \bibinfo{person}{Jake
  Cockrell}, \bibinfo{person}{William~G. Griswold}, {and}
  \bibinfo{person}{David Notkin}.} \bibinfo{year}{1999}\natexlab{}.
\newblock \showarticletitle{Dynamically Discovering Likely Program Invariants
  to Support Program Evolution}. In \bibinfo{booktitle}{\emph{Proc. of the 21st
  International Conference on Software Engineering}} (Los Angeles, California,
  USA) \emph{(\bibinfo{series}{ICSE '99})}. \bibinfo{publisher}{Association for
  Computing Machinery}, \bibinfo{pages}{213–224}.
\newblock
\showISBNx{1581130740}
\urldef\tempurl%
\url{https://doi.org/10.1145/302405.302467}
\showDOI{\tempurl}


\bibitem[Even-Mendoza et~al\mbox{.}(2022)]%
        {Karine22}
\bibfield{author}{\bibinfo{person}{Karine Even-Mendoza},
  \bibinfo{person}{Cristian Cadar}, {and} \bibinfo{person}{Alastair~F.
  Donaldson}.} \bibinfo{year}{2022}\natexlab{}.
\newblock \showarticletitle{CsmithEdge: More Effective Compiler Testing by
  Handling Undefined Behaviour Less Conservatively}.
\newblock \bibinfo{journal}{\emph{Empirical Softw. Engg.}}
  \bibinfo{volume}{27}, \bibinfo{number}{6} (\bibinfo{date}{nov}
  \bibinfo{year}{2022}), \bibinfo{numpages}{35}~pages.
\newblock
\showISSN{1382-3256}
\urldef\tempurl%
\url{https://doi.org/10.1007/s10664-022-10146-1}
\showDOI{\tempurl}


\bibitem[Fioraldi et~al\mbox{.}(2021)]%
        {Fioraldi-USENIX21}
\bibfield{author}{\bibinfo{person}{Andrea Fioraldi},
  \bibinfo{person}{Daniele~Cono D{\textquoteright}Elia}, {and}
  \bibinfo{person}{Davide Balzarotti}.} \bibinfo{year}{2021}\natexlab{}.
\newblock \showarticletitle{The Use of Likely Invariants as Feedback for
  Fuzzers}. In \bibinfo{booktitle}{\emph{30th {USENIX} Security Symposium
  ({USENIX} Security 21)}}. \bibinfo{publisher}{{USENIX} Association},
  \bibinfo{pages}{2829--2846}.
\newblock
\showISBNx{978-1-939133-24-3}
\urldef\tempurl%
\url{https://www.usenix.org/conference/usenixsecurity21/presentation/fioraldi}
\showURL{%
\tempurl}


\bibitem[gcc~bug tracker(2022a)]%
        {bug-gcc-104938}
\bibfield{author}{\bibinfo{person}{gcc~bug tracker}.}
  \bibinfo{year}{2022}\natexlab{a}.
\newblock \bibinfo{title}{{gcc bug ID: 104938}}.
\newblock
  \bibinfo{howpublished}{\url{https://gcc.gnu.org/bugzilla/show_bug.cgi?id=104938}}.
\newblock


\bibitem[gcc~bug tracker(2022b)]%
        {bug-gcc-105108}
\bibfield{author}{\bibinfo{person}{gcc~bug tracker}.}
  \bibinfo{year}{2022}\natexlab{b}.
\newblock \bibinfo{title}{{gcc bug ID: 105108}}.
\newblock
  \bibinfo{howpublished}{\url{https://gcc.gnu.org/bugzilla/show_bug.cgi?id=105108}}.
\newblock


\bibitem[gcc~bug tracker(2022c)]%
        {bug-gcc-105145}
\bibfield{author}{\bibinfo{person}{gcc~bug tracker}.}
  \bibinfo{year}{2022}\natexlab{c}.
\newblock \bibinfo{title}{{gcc bug ID: 105145}}.
\newblock
  \bibinfo{howpublished}{\url{https://gcc.gnu.org/bugzilla/show_bug.cgi?id=105145}}.
\newblock


\bibitem[gcc~bug tracker(2022d)]%
        {bug-gcc-105158}
\bibfield{author}{\bibinfo{person}{gcc~bug tracker}.}
  \bibinfo{year}{2022}\natexlab{d}.
\newblock \bibinfo{title}{{gcc bug ID: 105158}}.
\newblock
  \bibinfo{howpublished}{\url{https://gcc.gnu.org/bugzilla/show_bug.cgi?id=105158}}.
\newblock


\bibitem[gcc~bug tracker(2022e)]%
        {bug-gcc-105161}
\bibfield{author}{\bibinfo{person}{gcc~bug tracker}.}
  \bibinfo{year}{2022}\natexlab{e}.
\newblock \bibinfo{title}{{gcc bug ID: 105161}}.
\newblock
  \bibinfo{howpublished}{\url{https://gcc.gnu.org/bugzilla/show_bug.cgi?id=105161}}.
\newblock


\bibitem[gcc~bug tracker(2022f)]%
        {bug-gcc-105179}
\bibfield{author}{\bibinfo{person}{gcc~bug tracker}.}
  \bibinfo{year}{2022}\natexlab{f}.
\newblock \bibinfo{title}{{gcc bug ID: 105179}}.
\newblock
  \bibinfo{howpublished}{\url{https://gcc.gnu.org/bugzilla/show_bug.cgi?id=105179}}.
\newblock


\bibitem[gcc~bug tracker(2022g)]%
        {bug-gcc-105249}
\bibfield{author}{\bibinfo{person}{gcc~bug tracker}.}
  \bibinfo{year}{2022}\natexlab{g}.
\newblock \bibinfo{title}{{gcc bug ID: 105249}}.
\newblock
  \bibinfo{howpublished}{\url{https://gcc.gnu.org/bugzilla/show_bug.cgi?id=105249}}.
\newblock


\bibitem[gdb~bug tracker(2022)]%
        {gdb-28987}
\bibfield{author}{\bibinfo{person}{gdb~bug tracker}.}
  \bibinfo{year}{2022}\natexlab{}.
\newblock \bibinfo{title}{{gdb bug ID: 28987}}.
\newblock
  \bibinfo{howpublished}{\url{https://sourceware.org/bugzilla/show_bug.cgi?id=28987}}.
\newblock


\bibitem[Hennessy(1982)]%
        {Hennessy-TOPLAS82}
\bibfield{author}{\bibinfo{person}{John Hennessy}.}
  \bibinfo{year}{1982}\natexlab{}.
\newblock \showarticletitle{Symbolic Debugging of Optimized Code}.
\newblock \bibinfo{journal}{\emph{ACM Trans. Program. Lang. Syst.}}
  \bibinfo{volume}{4}, \bibinfo{number}{3} (\bibinfo{date}{jul}
  \bibinfo{year}{1982}), \bibinfo{pages}{323–344}.
\newblock
\showISSN{0164-0925}
\urldef\tempurl%
\url{https://doi.org/10.1145/357172.357173}
\showDOI{\tempurl}


\bibitem[Jaramillo et~al\mbox{.}(2000)]%
        {Jaramillo-SAS00}
\bibfield{author}{\bibinfo{person}{Clara Jaramillo}, \bibinfo{person}{Rajiv
  Gupta}, {and} \bibinfo{person}{Mary~Lou Soffa}.}
  \bibinfo{year}{2000}\natexlab{}.
\newblock \showarticletitle{FULLDOC: A Full Reporting Debugger for Optimized
  Code}. In \bibinfo{booktitle}{\emph{Proc. of the 7th International Symposium
  on Static Analysis}} \emph{(\bibinfo{series}{SAS '00})}.
  \bibinfo{publisher}{Springer-Verlag}, \bibinfo{pages}{240–259}.
\newblock
\showISBNx{3540676686}


\bibitem[Jia and Chan(2013)]%
        {Jia-AST13}
\bibfield{author}{\bibinfo{person}{Changjiang Jia} {and} \bibinfo{person}{W.~K.
  Chan}.} \bibinfo{year}{2013}\natexlab{}.
\newblock \showarticletitle{Which Compiler Optimization Options Should I Use
  for Detecting Data Races in Multithreaded Programs?}. In
  \bibinfo{booktitle}{\emph{Proc. of the 8th International Workshop on
  Automation of Software Test}} (San Francisco, California)
  \emph{(\bibinfo{series}{AST '13})}. \bibinfo{publisher}{IEEE Press},
  \bibinfo{pages}{53–56}.
\newblock
\showISBNx{9781467361613}


\bibitem[Leroy(2009)]%
        {CompCert-Com09}
\bibfield{author}{\bibinfo{person}{Xavier Leroy}.}
  \bibinfo{year}{2009}\natexlab{}.
\newblock \showarticletitle{Formal Verification of a Realistic Compiler}.
\newblock \bibinfo{journal}{\emph{Commun. ACM}} \bibinfo{volume}{52},
  \bibinfo{number}{7} (\bibinfo{date}{jul} \bibinfo{year}{2009}),
  \bibinfo{pages}{107–115}.
\newblock
\showISSN{0001-0782}
\urldef\tempurl%
\url{https://doi.org/10.1145/1538788.1538814}
\showDOI{\tempurl}


\bibitem[Li et~al\mbox{.}(2020)]%
        {Davide-PLDI20}
\bibfield{author}{\bibinfo{person}{Yuanbo Li}, \bibinfo{person}{Shuo Ding},
  \bibinfo{person}{Qirun Zhang}, {and} \bibinfo{person}{Davide Italiano}.}
  \bibinfo{year}{2020}\natexlab{}.
\newblock \showarticletitle{Debug Information Validation for Optimized Code}.
  In \bibinfo{booktitle}{\emph{Proc. of the 41st ACM SIGPLAN Conference on
  Programming Language Design and Implementation}} (London, UK)
  \emph{(\bibinfo{series}{PLDI 2020})}. \bibinfo{publisher}{Association for
  Computing Machinery}, \bibinfo{pages}{1052–1065}.
\newblock
\showISBNx{9781450376136}
\urldef\tempurl%
\url{https://doi.org/10.1145/3385412.3386020}
\showDOI{\tempurl}


\bibitem[{LLVM bug tracker}(2021a)]%
        {clang-49546}
\bibfield{author}{\bibinfo{person}{{LLVM bug tracker}}.}
  \bibinfo{year}{2021}\natexlab{a}.
\newblock \bibinfo{title}{{clang bug ID: 49546}}.
\newblock
  \bibinfo{howpublished}{\url{https://bugs.llvm.org/show_bug.cgi?id=49546}}.
\newblock


\bibitem[{LLVM bug tracker}(2021b)]%
        {bug-clang-49975}
\bibfield{author}{\bibinfo{person}{{LLVM bug tracker}}.}
  \bibinfo{year}{2021}\natexlab{b}.
\newblock \bibinfo{title}{{clang bug ID: 49975}}.
\newblock
  \bibinfo{howpublished}{\url{https://bugs.llvm.org/show_bug.cgi?id=49975}}.
\newblock


\bibitem[{LLVM bug tracker}(2021c)]%
        {clang-51780}
\bibfield{author}{\bibinfo{person}{{LLVM bug tracker}}.}
  \bibinfo{year}{2021}\natexlab{c}.
\newblock \bibinfo{title}{{clang bug ID: 51780}}.
\newblock
  \bibinfo{howpublished}{\url{https://bugs.llvm.org/show_bug.cgi?id=51780}}.
\newblock


\bibitem[{LLVM bug tracker}(2021d)]%
        {lldb-50076}
\bibfield{author}{\bibinfo{person}{{LLVM bug tracker}}.}
  \bibinfo{year}{2021}\natexlab{d}.
\newblock \bibinfo{title}{{lldb bug ID: 50076}}.
\newblock
  \bibinfo{howpublished}{\url{https://bugs.llvm.org/show_bug.cgi?id=50076}}.
\newblock


\bibitem[{LLVM bug tracker}(2022a)]%
        {bug-clang-53855}
\bibfield{author}{\bibinfo{person}{{LLVM bug tracker}}.}
  \bibinfo{year}{2022}\natexlab{a}.
\newblock \bibinfo{title}{{clang bug ID: 53855}}.
\newblock
  \bibinfo{howpublished}{\url{https://github.com/llvm/llvm-project/issues/53855}}.
\newblock


\bibitem[{LLVM bug tracker}(2022b)]%
        {report-llvm-arch64}
\bibfield{author}{\bibinfo{person}{{LLVM bug tracker}}.}
  \bibinfo{year}{2022}\natexlab{b}.
\newblock \bibinfo{title}{{clang bug ID: 54757}}.
\newblock
  \bibinfo{howpublished}{\url{https://github.com/llvm/llvm-project/issues/54757}}.
\newblock


\bibitem[{LLVM bug tracker}(2022c)]%
        {clang-55115}
\bibfield{author}{\bibinfo{person}{{LLVM bug tracker}}.}
  \bibinfo{year}{2022}\natexlab{c}.
\newblock \bibinfo{title}{{clang bug ID: 55115}}.
\newblock
  \bibinfo{howpublished}{\url{https://github.com/llvm/llvm-project/issues/55115}}.
\newblock


\bibitem[Regehr et~al\mbox{.}(2012)]%
        {CReduce-PLDI12}
\bibfield{author}{\bibinfo{person}{John Regehr}, \bibinfo{person}{Yang Chen},
  \bibinfo{person}{Pascal Cuoq}, \bibinfo{person}{Eric Eide},
  \bibinfo{person}{Chucky Ellison}, {and} \bibinfo{person}{Xuejun Yang}.}
  \bibinfo{year}{2012}\natexlab{}.
\newblock \showarticletitle{Test-Case Reduction for C Compiler Bugs}. In
  \bibinfo{booktitle}{\emph{Proc. of the 33rd ACM SIGPLAN Conference on
  Programming Language Design and Implementation}} (Beijing, China)
  \emph{(\bibinfo{series}{PLDI '12})}. \bibinfo{publisher}{Association for
  Computing Machinery}, \bibinfo{pages}{335–346}.
\newblock
\showISBNx{9781450312059}
\urldef\tempurl%
\url{https://doi.org/10.1145/2254064.2254104}
\showDOI{\tempurl}


\bibitem[Sahoo et~al\mbox{.}(2013)]%
        {Sahoo-ASPLOS13}
\bibfield{author}{\bibinfo{person}{Swarup~Kumar Sahoo}, \bibinfo{person}{John
  Criswell}, \bibinfo{person}{Chase Geigle}, {and} \bibinfo{person}{Vikram
  Adve}.} \bibinfo{year}{2013}\natexlab{}.
\newblock \showarticletitle{Using Likely Invariants for Automated Software
  Fault Localization}. In \bibinfo{booktitle}{\emph{Proc. of the Eighteenth
  International Conference on Architectural Support for Programming Languages
  and Operating Systems}} (Houston, Texas, USA) \emph{(\bibinfo{series}{ASPLOS
  '13})}. \bibinfo{publisher}{Association for Computing Machinery},
  \bibinfo{pages}{139–152}.
\newblock
\showISBNx{9781450318709}
\urldef\tempurl%
\url{https://doi.org/10.1145/2451116.2451131}
\showDOI{\tempurl}


\bibitem[Schuler et~al\mbox{.}(2009)]%
        {Schuler-ISSTA09}
\bibfield{author}{\bibinfo{person}{David Schuler}, \bibinfo{person}{Valentin
  Dallmeier}, {and} \bibinfo{person}{Andreas Zeller}.}
  \bibinfo{year}{2009}\natexlab{}.
\newblock \showarticletitle{Efficient Mutation Testing by Checking Invariant
  Violations}. In \bibinfo{booktitle}{\emph{Proc. of the Eighteenth
  International Symposium on Software Testing and Analysis}} (Chicago, IL, USA)
  \emph{(\bibinfo{series}{ISSTA '09})}. \bibinfo{publisher}{Association for
  Computing Machinery}, \bibinfo{pages}{69–80}.
\newblock
\showISBNx{9781605583389}
\urldef\tempurl%
\url{https://doi.org/10.1145/1572272.1572282}
\showDOI{\tempurl}


\bibitem[Sun et~al\mbox{.}(2016)]%
        {CompBugs-OOPSLA16}
\bibfield{author}{\bibinfo{person}{Chengnian Sun}, \bibinfo{person}{Vu Le},
  {and} \bibinfo{person}{Zhendong Su}.} \bibinfo{year}{2016}\natexlab{}.
\newblock \showarticletitle{Finding Compiler Bugs via Live Code Mutation}. In
  \bibinfo{booktitle}{\emph{Proc. of the 2016 ACM SIGPLAN International
  Conference on Object-Oriented Programming, Systems, Languages, and
  Applications}} (Amsterdam, Netherlands) \emph{(\bibinfo{series}{OOPSLA
  2016})}. \bibinfo{publisher}{Association for Computing Machinery},
  \bibinfo{pages}{849–863}.
\newblock
\showISBNx{9781450344449}
\urldef\tempurl%
\url{https://doi.org/10.1145/2983990.2984038}
\showDOI{\tempurl}


\bibitem[Wu et~al\mbox{.}(1999)]%
        {LeChun-PLDI99}
\bibfield{author}{\bibinfo{person}{Le-Chun Wu}, \bibinfo{person}{Rajiv Mirani},
  \bibinfo{person}{Harish Patil}, \bibinfo{person}{Bruce Olsen}, {and}
  \bibinfo{person}{Wen-mei~W. Hwu}.} \bibinfo{year}{1999}\natexlab{}.
\newblock \showarticletitle{A New Framework for Debugging Globally Optimized
  Code}. In \bibinfo{booktitle}{\emph{Proc. of the ACM SIGPLAN 1999 Conference
  on Programming Language Design and Implementation}} (Atlanta, Georgia, USA)
  \emph{(\bibinfo{series}{PLDI '99})}. \bibinfo{publisher}{Association for
  Computing Machinery}, \bibinfo{pages}{181–191}.
\newblock
\showISBNx{1581130945}
\urldef\tempurl%
\url{https://doi.org/10.1145/301618.301663}
\showDOI{\tempurl}


\bibitem[Yang et~al\mbox{.}(2011)]%
        {Yang-PLDI11}
\bibfield{author}{\bibinfo{person}{Xuejun Yang}, \bibinfo{person}{Yang Chen},
  \bibinfo{person}{Eric Eide}, {and} \bibinfo{person}{John Regehr}.}
  \bibinfo{year}{2011}\natexlab{}.
\newblock \showarticletitle{Finding and Understanding Bugs in {C} Compilers}.
  In \bibinfo{booktitle}{\emph{Proc. of the 32nd ACM SIGPLAN Conference on
  Programming Language Design and Implementation}} (San Jose, California, USA)
  \emph{(\bibinfo{series}{PLDI '11})}. \bibinfo{publisher}{Association for
  Computing Machinery}, \bibinfo{pages}{283–294}.
\newblock
\showISBNx{9781450306638}
\urldef\tempurl%
\url{https://doi.org/10.1145/1993498.1993532}
\showDOI{\tempurl}


\end{thebibliography}

\appendix

% !TEX root = main.tex

\section{Appendix}
\label{se:appendix}

In this online appendix, we provide a brief analysis of each of the 38 issues that we reported to the developers of the clang/lldb and gcc/gdb ecosystems. Issues are listed in their order of appearance in Table~\ref{tab:reports} along with the conjecture that exposed it and, if applicable, the category of DWARF-level manifestation we noticed for them.

\subsection*{clang}

\paragraph{49546 (C1, Missing DIE)\label{rep:49546}} Described in Section~\ref{ss:eval-bugs}. An induction variable is not available during debugging when passed as call argument to an opaque function. Loop optimizations can figure out that only a single iteration will take place and the two possible values of the variable should be handled using a location attribute with two ranges. Due to implementation defects, debug information for either is lost by distinct LLVM passes. Only {\tt -Og} is affected.

\paragraph{49580 (C1, Missing DIE)\label{rep:49580}} An induction variable is not available during debugging when passed as call argument to an opaque function. Loop rotation does not push debug metadata that LLVM emits for variable values to the exit block of the loop. Then, when loop reduction optimizes the loop away, the debug metadata information ends up being attached to an undefined location. Eventually, debug information is not emitted at all in DWARF. Only {\tt -Og} is affected.

\paragraph{49769 (C1, Hollow DIE)\label{rep:49769}} Several constant-value variables are not available during debugging when passed as call arguments to an opaque function. The control-flow graph simplification that follows inlining results in the removal of IR-level debug statements when these are the only elements in a block. Only {\tt -Og} is affected.

\paragraph{49973 (C1, Hollow DIE)\label{rep:49973}} A constant-value variable is not available during debugging when passed as argument to an opaque function called within a loop. This is caused by the simplification of induction variables, which results in not propagating the debug information about the constant value. Only {\tt -O3} is affected. The analysis of this issue also led us to discovering an lldb bug (bug 50076).

\paragraph{49975 (C1, Hollow DIE)\label{rep:49975}} Described in Section~\ref{ss:invariant-one}. Local variable {\tt l\_28} is not available during debugging when passed as call argument to an opaque function. This is caused by the peephole optimization of the bitwise AND in the expression {\tt l\_69 = (l\_28 = a) == 0 \& c;} where {\tt l\_28} is first assigned. A developer noted that debug information could have been left attached to the {\tt select} statement introduced after the simplification. Only {\tt -O3} is affected.

\paragraph{51780 (C1, Missing DIE)\label{rep:51780}} A local variable assigned by reading a global variable is not available during debugging when passed as call argument to an opaque function. This occurs during instruction selection due to an acknowledged gap in the LLVM infrastructure for retaining debug information for this case. Only {\tt -O2} is affected.

\paragraph{55101 (C1, Hollow DIE)\label{rep:55101}} A local variable is not available during debugging when passed as argument to an opaque function called within a loop. Two optimizations cause a progressive loss of debug information. First, loop strength reduction drops location information for the instructions belonging to the loop only (therefore, the variable would appear as optimized out there and available elsewhere). Then, instruction selection causes the loss of the information on the other locations as well. Only {\tt -O2} is affected.

\paragraph{55115 (C1, Missing DIE)\label{rep:55115}} A local variable is not available during debugging when passed as argument to an opaque function called within a loop. Similarly to 49769, the issue is caused by control-flow graph simplification removing debug statements at IR level, due to the unfeasibility (acknowledged by developers) to put it anywhere else in the IR. {\tt -O1}, {\tt -O2}, {\tt -O3}, and {\tt -Og} are affected.

\paragraph{55123 (C1, Hollow DIE)\label{rep:55123}} A constant-value variable is not available during debugging when passed as argument to an opaque function called within a loop. The issue is caused by an interaction of two optimizations (instruction combining and inlining) that wrongly update the location in debug statements at IR level. In particular, instruction combining associates debug metadata in the IR with an undefined location. {\tt -O1}, {\tt -O2}, {\tt -O3}, and {\tt -Og} are affected.

\paragraph{53855a (C2, Hollow DIE)\label{rep:53855a}} Described in Section~\ref{ss:invariant-three}. An induction variable is not available when used in an expression for assigning to global storage within a loop. This was caused by the loop strength reduction pass not correctly salvaging the debug statements regarding the variable. The issue was independently fixed by developers in version {\tt trunk*}. {\tt -O1}, {\tt -Og}, and {\tt -Oz} were affected.

\paragraph{53855b (C2, Hollow DIE)\label{rep:53855b}} Described in Section~\ref{ss:invariant-three}. Similarly to 53855a above, an induction variable is not available when used in another expression for assigning to global storage within a loop. LSR is again the culprit optimization but, unlike 53855a, the issue is not handled by the fix introduced in {\tt trunk*}. Only {\tt -Os} is affected.

\paragraph{54611 (C2, Incomplete DIE)\label{rep:54611}} A local variable is not available when used in an expression for assigning to global storage. This is caused by instruction scheduling leading to an incomplete range definition, which no longer includes the (new) instruction associated with the source line assigning the expression. Only {\tt -O1} is affected.

\paragraph{54757 (C2, Hollow DIE)\label{rep:54757}} An induction variable is not available when used in an expression assigned to global storage within a loop. This is caused by loop optimizations removing the loop and, when doing so, also dropping part of debug information associated with the expression. In particular, the output code contains a single instruction for the assignment {\tt a = (i)*j;}, where {\tt i} is available while {\tt j} is optimized out. {\tt -O1}, {\tt -O2}, {\tt -O3} and {\tt -Og} are affected. Although the optimization pipeline and transformations are completely different, we saw that recent gcc versions at {\tt -O2} emit identical code and complete debugging information, using {\tt DW\_at\_location} in the DIE for {\tt j} to distinguish two ranges for its values {\tt 0} and {\tt 1}.

\paragraph{54763 (C2, Incomplete DIE)\label{rep:54763}} Two constant-value variables are not available when used in an expression assigned to global storage. Apparently, this may be caused by the unfeasibility to put debug statements in a block before $\phi$-nodes in LLVM IR. The variables eventually become available after the corresponding source-level assignment, which is done using $\phi$-nodes. {\tt -O2} and {\tt -O3} are affected.

\paragraph{50286 (C3, Incomplete DIE)\label{rep:50286}} A local variable presents intermittent availability during its lifetime. An issue in instruction scheduling causes the debug information location to not include all the assembly instructions associated with the source lines where the variable should be available. Only {\tt -Og} is affected.

\paragraph{54796 (C3, Incomplete DIE)\label{rep:54796}} A local variable presents intermittent availability during its lifetime. The scalar replacement of aggregates (SROA) optimization of LLVM removes the location attribute from the debug information and later control-flow graph simplification restores it, but only partially. Only {\tt -Os} is affected.

\subsection*{gcc}

\paragraph{104549 (C1, Incorrect DIE)\label{rep:104549}} A constant-value variable is not available during debugging when passed as call argument to an opaque function. Inlining wrongly updates the location definition of the function where the variable is live, despite the optimizer kept track of the value. In our best judgement, the initial reaction from the developer---asking for clarification on the reported issue, claiming that the information was complete---came from analyzing the issue only statically, i.e., they overlooked what DIE the debugger actually retrieves when stepping on the statement. {\tt -O2} and {\tt -O3} are affected.

\paragraph{105007 (C1, Hollow DIE)\label{rep:105007}} Two local variables are not available during debugging when passed as call arguments to an opaque function. Their location definitions are missing in DWARF: a thorough analysis conducted by the developers revealed that the EVRP lattice propagation did not insert a debug statement when removing a definition for a propagated constant. {\tt -O2} and {\tt -O3} are affected.

\paragraph{105158 (C1, Hollow DIE)\label{rep:105158}} A local variable is not available during debugging when passed as call argument to an opaque function. This is caused by an issue in the cleanup of the control-flow graph performed after optimizing the boolean expression assigning the variable. A developer patched the bug in few days. As discussed in Section~\ref{ss:eval-regression}, since the cleanup is done upon other optimizations too, the enhancement brought general improvements to variable availability in our tests. {\tt -O1}, {\tt -O2}, {\tt -O3}, and {\tt -Og} are affected.

\paragraph{105176 (C1, Incomplete DIE)\label{rep:105176}} A local variable is not available during debugging when passed as argument to an opaque function called within a loop. In our analysis, dead code elimination drops debug information without changing the code produced by the compiler for the test program. {\tt -Os} and {\tt -Oz} are affected.

\paragraph{105179 (C1, Incomplete DIE)\label{rep:105179}} Described in Section~\ref{ss:eval-bugs}. A local variable is not available during debugging when passed as argument to an opaque function called within a loop. Transformation {\tt -fcprop-registers} attempts to reduce scheduling dependencies but, when active, leads gcc to emit a range for the variable that does not include the address of the call. Only {\tt -Og} is affected.

\paragraph{105239 (C1, Incomplete DIE)\label{rep:105239}} A local variable is not available during debugging when passed as call argument to an opaque function called within a loop. Only {\tt -Og} is affected. The test case hosts a call to a function taking no arguments right before the incriminated call. Our DWARF analysis revealed that the location definition in the DIE of the variable does not include the address of the call to the opaque function. If the other call is commented out, though, the variable becomes available (and variable DIE data becomes identical to what {\tt -O1} would produce for the original program). Although the issue is not yet confirmed as a bug, a developer commented that there is a difference materializing somewhere on the RTL side, which is the low-level IR for the gcc infrastructure.
 
\paragraph{105248 (C1, Hollow DIE)\label{rep:105248}} A local variable is not available during debugging when passed as call argument to an opaque function. Similarly to 105176, the loss is caused by the dead store elimination transformation dropping debug information without changing the output code. {\tt -O1}, {\tt -O2}, and {\tt -O3} are affected.

\paragraph{105261 (C1, Hollow DIE)\label{rep:105261}} Several constant-value variables are not available during debugging when passed as call arguments to an opaque function. This is caused by the scalar replacement of aggregates optimization and, depending on the chosen optimization level, by its interaction with instruction scheduling. From this test case, we also filed 29060 to gdb developers, since also variables for which debug information is correctly defined are not visible at the same program point. {\tt -O2}, {\tt -O3}, {\tt -Os}, and {\tt -Oz} are affected.

\paragraph{104891  (C2, Incomplete DIE)\label{rep:104891}} Several constant-value variables are not available when used in an expression assigned to global storage. The issue is caused by incomplete location definitions and is only present when the variable declarations and the involved expression are located within an unnamed scope, i.e., within brackets. Even if the brackets do not change the semantic of this program, as constructs they may still induce optimizations to harm debug information during its updating. {\tt -O2} and {\tt -O3} are affected.

\paragraph{105036 (C2, Incorrect DIE)\label{rep:105036}} An induction variable is not available when used in an expression assigned to global storage within a loop. Incorrect program point information causes the debugger to display a wrong current function frame, leading to the variable not being available. The issue is a combined effect of instruction scheduling, function inlining, and loop unrolling. Only {\tt -O3} is affected.

\paragraph{105108 (C2, Hollow DIE)\label{rep:105108}} Described in Section~\ref{ss:eval-bugs}. A constant-value variable is not available when used in an expression assigned to global storage. The issue is caused by a combination of constant propagation and value range propagation exposing an infrastructural limitation of gcc. The discussion between developers hinted to an extension to the DWARF 6 format. {\tt -Og} and {\tt -O1} are affected.

\paragraph{105145 (C2, Hollow DIE)\label{rep:105145}} A pointer-type variable is not available when used in an expression assigned to global storage. As a developer observed, the issue is caused by the current inability of gcc to retain debug information for address-taken local variables that eventually become registers later on. Some gcc developer(s) tried to work on this limitation in the past. {\tt -O1}, {\tt -O2}, and {\tt -O3} are affected.

\paragraph{105161 (C2, Hollow DIE)\label{rep:105161}} Described in Section~\ref{se:intro}. A constant-value variable is not available when used in an expression assigned to global storage within a loop. Constant folding is behind the issue. The bug got the attention of the developers as it is similar to 105158 but not straightforward to fix. A technically rich discussion outlined several angles to explore. {\tt -O1}, {\tt -O2}, {\tt -O3}, and {\tt -Og} are affected.

\paragraph{105249 (C2, Incorrect DIE)\label{rep:105249}} Described in Section~\ref{ss:eval-bugs}. An induction variable is not available when used in an expression assigned to global storage. Similarly to 105036, incorrect program point information causes the wrong function frame to be displayed as current, leading to the variable not being available. Differently from 105036, this happens only due to instruction scheduling, since the structure of the inlined function is much simpler here. Only {\tt -Os} is affected.

\paragraph{104938 (C3, Incomplete DIE)\label{rep:104938}} Described in Section~\ref{ss:invariant-two}. A local variable presents intermittent availability during its lifetime. Conditional constant propagation is responsible for shrinking the range in the location definition of the variable. Only {\tt -Og} is affected.

\paragraph{105124 (C3, Incomplete DIE)\label{rep:105124}} A constant-value variable presents intermittent availability during its lifetime. The location definition of the variable does not include some instructions associated with source lines where the variable is live. Interestingly, availability may be influenced by the value chosen for it. Only {\tt -Og} is affected.

\paragraph{105159 (C3, Hollow DIE)\label{rep:105159}} A local variable presents intermittent availability during its lifetime. The issue may be caused by {\tt -fipa -reference-addressable}, which discovers read-only, write-only, and non-addressable static variables. The location definition of the variable is lost while the code stays the same. Only {\tt -Og} is affected.

\paragraph{105194 (C3, Incomplete DIE)\label{rep:105194}} A local variable presents intermittent availability during its lifetime. After dead code elimination takes place, cleaning up the control-flow graph wrongly updates the location definition of the variable. This bug was fixed by the patch developed for 105158. {\tt -O1}, {\tt -O2}, {\tt -O3}, and {\tt -Og} were affected.

\paragraph{105389 (C3, Incomplete DIE)\label{rep:105389}} A local variable presents intermittent availability during its lifetime. Its location is defined using a range for every value the variable assumes during its lifetime, but one value is missing causing a range of addresses not to be covered by debug symbols. Only {\tt -Og} is affected. Incidentally, the test case also exposed a correctness bug on the same variable with {\tt -O2}.

\vspace{-2pt}
\subsection*{gdb}

\paragraph{28987 (C1)\label{rep:28987}} We spotted this bug when triaging the culprit optimization for 105007: when disabling certain optimizations, gdb ends up displaying an outdated value for an induction variable passed as call argument to an opaque function. The location of the variable is defined with a sequence of ranges, some of which having identical low and high addresses. lldb could handle such ranges correctly.

\paragraph{29060 (C1)\label{rep:29060}} \looseness=-1 Some variables are not available during debugging when passed as call arguments to an opaque function. The enclosing function is inlined into {\tt main()}. In the DWARF information we found in the program, the concrete and the abstract representation of the inlined function differ in structure. In particular, the former contains a lexical block that the latter does not. Unlike lldb, this discrepancy makes gdb unable to display the values for said variables.

\vspace{-2pt}
\subsection*{lldb}

\paragraph{50076 (C1)\label{rep:50076}} Described in Section~\ref{ss:eval-bugs}. A constant-value variable is not available in lldb when passed as call argument to an opaque function called within a loop. In the DWARF information for this program, the location of the variable is only defined in the abstract origin of the inlined function where it is live. Since the concrete representation of said function does not include a location definition for the variable, lldb is unable to display its value, while gdb can.

Interestingly, the bug above and 29060 for gdb involve symmetric discrepancies between abstract and concrete representations of inlined functions: as far as we know, both are legitimate in DWARF.

\end{document}